\documentclass[useAMS,usenatbib]{mn2e}

\usepackage{aas_macros}
\usepackage{dcolumn}
\usepackage{amsmath}
\usepackage{amssymb}
\usepackage{graphicx}
\usepackage{longtable}
\newcommand\refcom[1]{{#1}}%

\newcommand{\bagatelle}{\textsc{bagatelle}}
\newcommand{\sex}{\textsc{sextractor}}

\def\Reff{\ifmmode{R_\mathrm{eff}}\else{$R_\mathrm{eff}$}\fi}
\def\mueff{\ifmmode{\mu_\mathrm{eff}}\else{$\mu_\mathrm{eff}$}\fi}

\begin{document}
\title[Colour gradients in Coma galaxies]{The HST/ACS Coma Cluster Survey: VII - Colour Gradients in Giant and Dwarf Early-Type Galaxies}
\author[M. den Brok et al.]{M.~den~Brok,$^1$\thanks{denbrok@astro.rug.nl} R.F.~Peletier,$^1$ E.A.~Valentyn,$^1$ M.~Balcells,$^{2,3,4}$,D.~Carter,$^5$  P.~Erwin,$^{6,7}$
\newauthor  H.C.~Ferguson,$^8$ P.~Goudfrooij,$^8$ A.W.~Graham,$^9$ D.~Hammer,$^{8,10}$ J.R.~Lucey,$^{11}$  
\newauthor
N.~Trentham,$^{12}$ R.~Guzm\'an,$^{13}$ C.~Hoyos,$^{14}$ G.~Verdoes Kleijn,$^1$ S.~Jogee,$^{15}$
\newauthor 
A.M.~Karick,$^5$ I.~Marinova,$^{15}$ M.~Mouhcine,$^{5}$ and  T.~Weinzirl$^{15}$ \\
$^1$Kapteyn Astronomical Institute, University of Groningen, P.O. Box 800, 9700AV Groningen, The Netherlands.\\
$^2$Isaac Newton Group of Telescopes, Apartado 321, 38700 Santa Cruz de La Palma, Canary Islands, Spain\\
$^3$Instituto de Astrof\'{\i}sica de Canarias, C. V\'{\i}a L\'{a}ctea, S/N, 38200 La Laguna, Tenerife, Spain\\
$^4$Departamento de Astrof\'{\i}sica, Universidad de La Laguna, 38200 La Laguna, Tenerife, Spain\\
$^5$Astrophysics Research Institute, Liverpool John Moores University, Twelve Quays House, Egerton Wharf, Birkenhead, CH41 1LD, UK.\\
$^6$Max-Planck-Institut f\"ur extraterrestrische Physik, Giessenbachstrasse, D-85748 Garching, Germany.\\
$^7$Universit\"ats-Sternwarte M\"unchen, Scheinerstrasse 1, D-81679 M\"unchen, Germany.\\
$^8$Space Telescope Science Institute, 3700 San Martin Drive, Baltimore, MD 21218, USA.\\
$^9$Centre for Astrophysics and Supercomputing, Swinburne University of Technology, Hawthorn, Victoria 3122, Australia.\\
$^{10}$Laboratory for X-Ray Astrophysics, NASA Goddard Space Flight Center, Code 662.0, Greenbelt, MD 20771, USA\\
$^{11}$Department of Physics, University of Durham, Durham DH1 3LE, UK.\\
$^{12}$Institute of Astronomy, Madingley Road, Cambridge CB3 0HA, UK.\\
$^{13}$Department of Astronomy, University of Florida, P.O. Box 112055, Gainesville, FL 32611, USA.\\
$^{14}$School of Physics and Astronomy, The University of Nottingham, University Park, Nottingham, NG7 2RD, UK.\\
$^{15}$Department of Astronomy, University of Texas at Austin, Austin, TX, USA.\\
}
\maketitle
\begin{abstract}
Using deep, high-spatial resolution imaging from the HST ACS Coma Cluster Treasury Survey, we determine colour profiles of early-type galaxies in the Coma cluster. 
From 176 galaxies brighter than $M_\mathrm{F814W(AB)} = -15$ mag that are either spectroscopically confirmed members of Coma or identified 
by eye as likely members from their low surface brightness, data are provided for 142 early-type galaxies. 
Typically, colour profiles are linear against $\log(R)$, sometimes with a nuclear region of distinct, often bluer colour associated with nuclear clusters.  Colour gradients are determined for the regions outside the nuclear components.  
We find that almost all colour gradients are negative, both for elliptical and lenticular galaxies.  Most likely, earlier studies that report positive colour gradients in dwarf galaxies are affected by the bluer colours of the nuclear clusters, underlining that high resolution data are essential to disentangle the colour properties of the different morphological components in galaxies.
Colour gradients of dwarf galaxies form a continuous sequence with those of elliptical galaxies, becoming shallower toward fainter magnitudes. 
Interpreting the colours as metallicity tracers, our data suggest that dwarfs as well as giant early-type galaxies in the Coma cluster are less metal rich in their outer parts. 
We do not find evidence for environmental influence on the gradients, although we note that most of our galaxies are found in the central regions of the cluster. For a subset of galaxies with known morphological types, S0 galaxies have less steep gradients than elliptical galaxies. 
\end{abstract}
\begin{keywords}
galaxies: elliptical and lenticular, cD -- galaxies: dwarf -- galaxies: clusters: individual: coma -- galaxies: structure
\end{keywords}
\section{Introduction}
Metallicity gradients provide a means of studying galaxy formation. Classical monolithic collapse scenarios \citep{Lar74,Car84,AriYos87} predict strong metallicity gradients. In this scenario primordial clouds of gas sink to the centre of an overdensity where a rapid burst of star formation occurs. Infalling gas mixes with enriched material freed from stars by stellar evolutionary processes and forms a more metal rich population. Because the gas clearing time (and hence the number of generations which enrich the interstellar medium) is dependent on the depth of the potential well, the metallicity gradient is dependent on the mass of the galaxy.

Mergers, dominant in hierarchical galaxy formation scenarios, will dilute existing population gradients \citep[e.g.][]{Whi80,diMPipLeh09}, \refcom{although residual central star formation can steepen gradients again \citep[e.g.][among others]{HopCoxDut09}} . Therefore, the study of metallicity gradients can be used to distinguish between competing scenarios of galaxy formation and can eventually lead to a more detailed understanding of those scenarios. 
Many attempts have been made to model metallicity gradients in more detail \citep{Car84,AriYos87,ChiCar02,Kob04,PipDErMat08,PipDErChi10}, of which the models by \citet{KawGib03} are so far most consistent with observations \citep{SpoProFor09}.

Much work has been devoted to determine metallicity gradients in large elliptical galaxies, either by use of colour gradients \citep{San72,FraIllHec89,PelDavIll90,PelValJam90,SagMarGre00,LaBMerBus04,LaBdeCGal05,LaBdeC09} or by spectroscopy \citep{MehThoSag03,SanGorCar06,OgaMaiPel08,SpoKobFor10}. There is consensus among all authors that elliptical galaxies do have metallicity gradients, which are not as strong as predicted by classic monolithic collapse scenarios.  
However, for dwarf galaxies, the situation is not as clear. \citet{VadVigLac88} find that colour gradients become positive for dwarf galaxies. \citet{vanBarSki04} mitigate this statement by saying that non-nucleated dwarf galaxies have positive gradients, but nucleated dwarf galaxies have more or less flat gradients. \citet{GorPedGuz97} find negative metallicity gradients for a sample of high-luminosity dwarf galaxies. The metallicity gradients of \citet{SpoProFor09} show a trend to flat or positive gradients at low luminosities, whereas the metallicity gradients from \citet{KoldeRPru09} stay again negative in this regime.

In addition to the different formation scenarios, different evolutionary paths may have played a r\^ole in shaping the dwarf galaxies in different parts of the cluster. It is obvious that at some point dEs have started as entities containing gas, however, as we observe them now they are gas-poor. The morphology-density relation \citep{Dre80} suggests that cluster-specific processes are important for removing the gas. Several different processes have been proposed to operate in clusters (and less dense environments), but it is not clear which one is the most important in the end. Ram pressure stripping by the hot ICM \citep{GunGot72} will result in a different stellar population than for example harassment \citep{MooKatLak96} and starvation \citep{TulTre08}. By looking at gradients in different environments it may be possible to identify the most prominent evolutionary paths (if such a thing exists, see for example \citep{vanBarSki04}.) It is obvious that this is extremely difficult, because all gas removing processes act more efficiently on gas in the outskirts of the galaxies and are more inefficient in the removal of gas in the centre. If the left-over gas in the centre forms stars, this will steepen existing gradients a bit, but it is not clear by how much (except in the case of tidal stripping, which removes also part of the stellar content of a galaxy). However, it is clear that these processes depend on galaxy density or position in the cluster. The Coma cluster is therefore an ideal laboratory to study the evolution of galaxies, as the density is to a first approximation only a function of clustercentric distance.  

Coma, at a distance of 100 Mpc the nearest Abell class 3 cluster, is an excellent place to study dwarf galaxies for various reasons. Since Coma is located at high Galactic latitude, the extinction is low \citep{SchFinDav98}. Furthermore, Coma, serving as a benchmark for high redshift studies, has been looked at intensively in the past -- both with ground based telescopes and with space-based telescopes (GALEX, Spitzer) -- and in the present. With the advent of the Coma Treasury survey many new observing programs have commenced.\footnote{An overview of observing programs can be found at \texttt{http://astronomy.swin.edu.au/coma}}

In this paper we study colour gradients for dwarf and giant galaxies in the Coma cluster as derived from HST data. Colours probe stellar populations: colours of metal rich stars become redder, but also the age of the population and the slope of the initial mass function (IMF) change the colour. The superior resolution of HST allows us to separate different morphological components of galaxies, leading to a clean interpretation of our observations. 

Throughout this paper we assume a distance modulus to Coma of $m-M=35$ \citep[henceforth paper I]{CarGouMob08}, which corresponds to a distance of 100Mpc. 

\section{Data and Observations}
\subsection{HST/ACS Coma Cluster Survey data}
\begin{figure*}
\begin{minipage}{182mm}
\center
\scalebox{0.45}[0.45]{\includegraphics{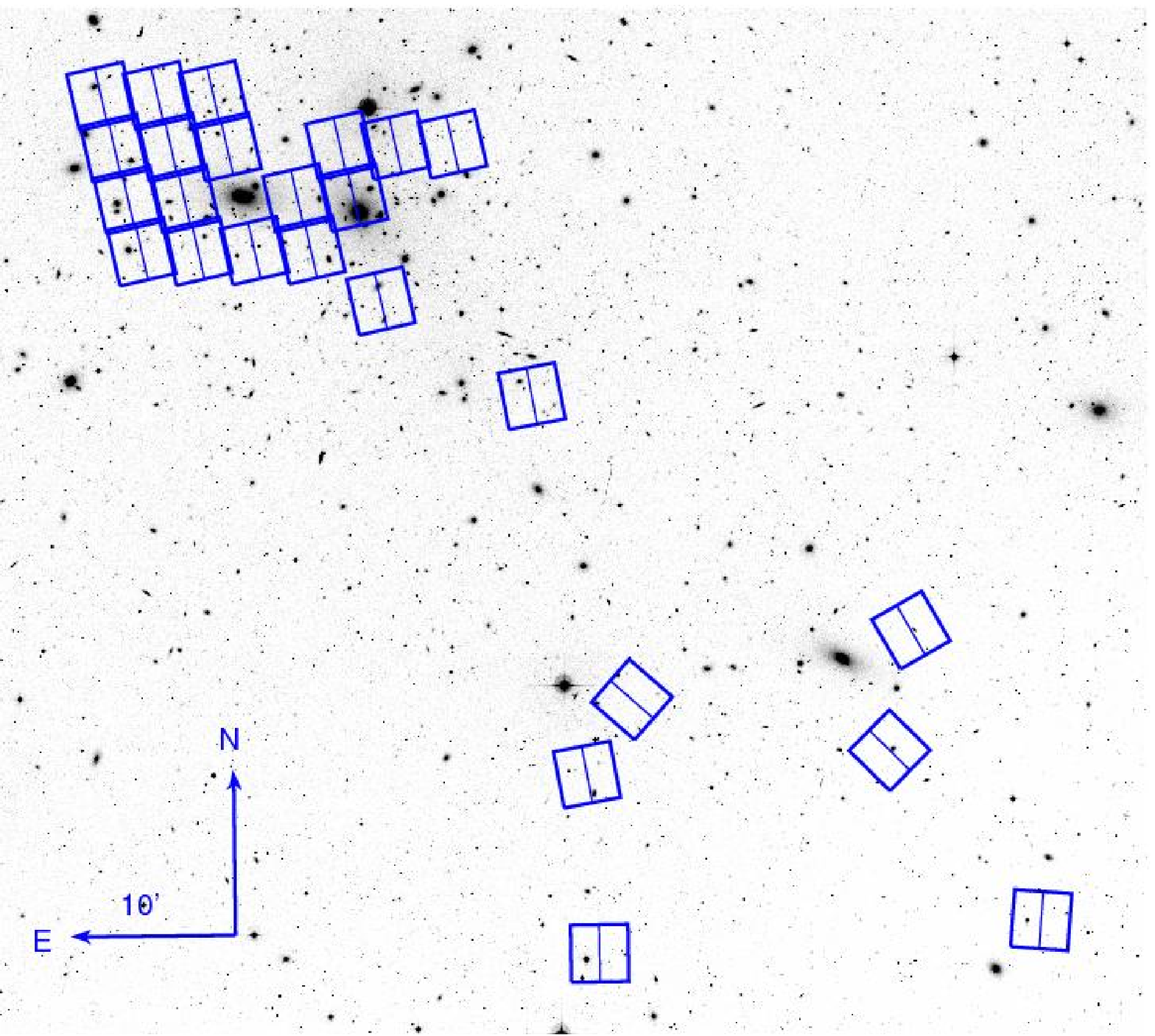}}
\scalebox{0.45}[0.45]{\includegraphics{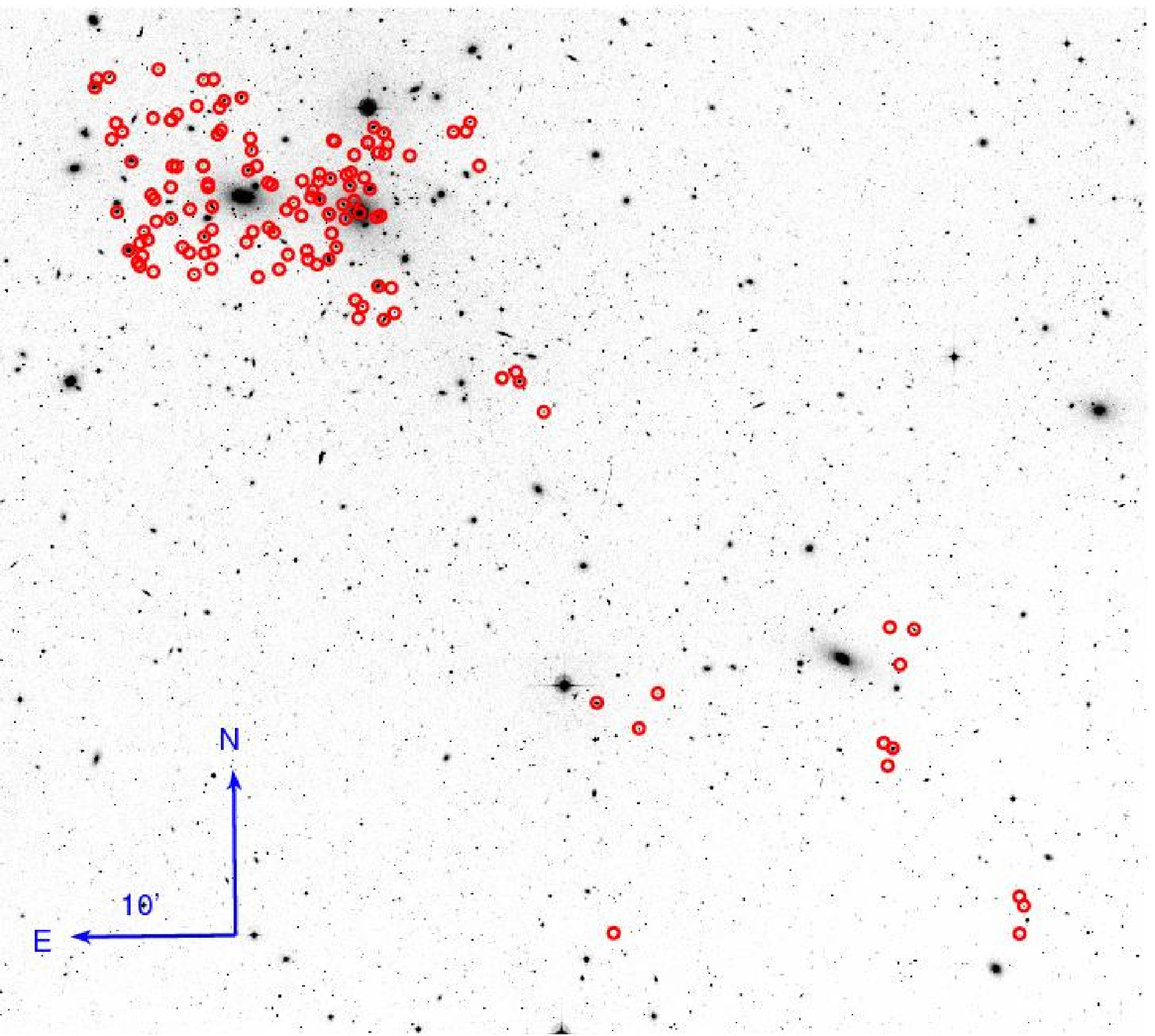}}
\caption{Left: Positions of the Coma ACS fields, overlayed on an XDSS image. The length of the arrow of the compass is $10^\prime$. Right: position of sample galaxies.}\label{fig:coma_pos}
\end{minipage}
\end{figure*}

We have used the data from the HST/ACS Coma Cluster Survey, a deep 2-passband imaging survey of the Coma Cluster. A full description of the observations and data reduction can be found in  Paper I. Here we will give only a short summary of the survey.
The ACS Wide Field Camera was used to image the Coma Cluster in the F814W and F475W passbands (roughly coinciding with the I and g filters). The survey was designed to contain different galaxy density regimes, by observing both the core of the cluster and the outskirts, and in particular the infalling group of NGC 4839. Due to the breakdown of the ACS Camera the survey is only partly completed; with 19/25 observed fields the centre of the cluster is well-covered. The other six fields sample the region around the infalling group NGC4839. Figure \ref{fig:coma_pos} shows the positions of the frames on the sky.
The fields that were completely observed (4/25 fields are only partly observed) had exposure times of $4\times350$s in F814W and $4\times640$s in F475W. After data reduction, which included PyRAF/STSDAS and MultiDrizzle (see paper I for details), each ACS frame consists of 4300$\times$4225 pixels, each having a size of 0.05''$\times0.05$''.\\
A source catalogue was created by running SExtractor on the reduced frames. A full description of the source catalogue is given in paper II \citep{HamVerHoy10}

\subsection{Sample selection}
Except for the galaxies in the most central fields, the majority of observed galaxies are background galaxies that do not belong to the cluster. For our analysis we are only interested in galaxies inside the Coma Cluster. 
Arguably the most reliable way of determining cluster membership is by determining the redshift of each galaxy. For a large sample of galaxies this has been done using the Hectospec spectograph on MMT, and the LRIS spectrograph on Keck (see \citet{SmiMarHor08,SmiLucHud09,ChiTulMar10}.)
The sample for which we have accurate redshift information is approximately complete for galaxies with $\mu_e$(F814W) $<$ 22.5.

The majority of the fainter Coma dwarf galaxies stand out against background galaxies as low surface brightness galaxies. Due to the superior resolution and sensitivity of the ACS camera it is therefore possible to estimate cluster membership solely on the basis of morphology for faint galaxies of which membership cannot be confirmed spectroscopically. We use the same sample as used for the determination of the luminosity function for Coma. For details on the selection method and the selected galaxies, see Trentham et al. (2010, in preparation). The sources selected in Trentham et al. all have a membership rating, going from 0 to 3. Class 0 sources are spectroscopically confirmed members of Coma. Class 1 contains ``almost certain'' members, which are sources that are only rarely found in the field. Class 2 contains ``likely'' members, and class 3 contains possible Coma members. Spectroscopic follow-up has shown that rating class 2 contains approximately 10\% non-members, whereas rating class 3 contains $\sim$50\% non-members. To minimise background contamination, we only use members from class 0 (confirmed) to class 2 (likely member), but leave out class 3. Our sample contains approximately 176 galaxies, 10\% of which are of class 2. This means that the expected number of non-members in our sample is 2. 

Not all galaxies in this sample can be used for determining colour gradients. First of all, we are only interested in early-type galaxies, and we therefore exclude galaxies with obvious spiral features. For sources fainter than F814W(AB) $=$ 20 the dwarfs become generally too faint to obtain any reliable results at all, so we also have excluded these sources. In addition to this, the sample contained a number of extremely compact galaxies (UCDs) which were picked up by Hectospec and LRIS. Since their size is hardly larger than the size of the PSF, we cannot obtain any useful colour gradients from them. We have excluded galaxies where the presence of dust in the outer parts of the galaxies prevented us from probing stellar population differences. In addition to this, there are a number of galaxies for which it is impossible to obtain reliable results, because they reside too close to other galaxies, and it is impossible to get a reliable estimate of the sky, or because the shape of the galaxy is such (highly flattened, or irregular) that the fitting procedure does not give very accurate results. Table~\ref{table:sample} in the Appendix contains the list of 142 fitted sources and their membership rating. Table~\ref{table:excluded} contains a list of 36 galaxies that have been excluded for various reasons. \refcom{Out of these 142 sources, 1 source has membership class 1, and 16 have membership class 2, meaning that we have cluster membership confirmation for 125 out of 142 sources ($88\%$).} 

\section{Analysis}
\subsection{Galaxy isophotal analysis}
We use Galphot \citep{FraIllHec89,JrgFraKjr92} to measure the surface brightness profile of each galaxy. This program fits ellipses to a light distribution, allowing a variable centre, ellipticity and position angle as function of radius. 

For our colour gradients, all the fitted ellipses need to have a common centre. The centre as determined by Sextractor is in many cases not accurate enough, because its luminosity weighted determination will not coincide with the real centre, for example, if a galaxy is slightly asymmetric or if the sky background has some gradient. In this case one can either choose to centre the ellipses, after an initial Galphot run, on the outer isophotes or on the inner isophotes, or some combination of both. Motivated by the presence of nuclei in most dwarf ellipticals, we decide to fix the centre on the inner ellipse. We note however that a slightly different centre has only a negligible effect on the gradient. 

The shape of the central light profile of a galaxy is usually quite sensitive to the size of the point spread function (PSF). To obtain useful colour profiles one has to compare fluxes at the same place in the galaxies. We convolved each image stamp in each band with the psf of the other bands. PSFs were generated (cf. \citet{HoydenVer10}, henceforth Paper III) using DrizzlyTim\footnote{DrizzlyTim is written by Luc Simard}. Convolution was done in frequency space using a Fast Fourier Transform. 

In summary, we set up our Galphot runs in the following way. In the first run, we give Galphot as input the centre of the source as determined by SExtractor, and allow the centres of ellipses to change. This Galphot run is done on a convolved F814W image stamp. In the second run, we use the results from the first run to fix the centre, ellipticy and position angle, such that they are the same for all ellipses. After this, we determine the azimutally averaged surface brightness of the same galaxy in the F475W band, using the same set of ellipses. A colour gradient is then determined by fitting a straight line in colour-log(r) space, where $r$ is the circularized distance of the ellipse (i.e.  the geometric mean of the major and minor axis.) This scale-free way of fitting avoids the use of a predetermined scale in a galaxy, such as the effective radius, and is common in the literature. Particularly for galaxies with strong gradients using a logaritmic gradient gives a better fit than a linear gradient. The range over which we fit extends from the innermost point to the cutoff in the profile (see section \ref{skydetermination}), and the points are weighted in a standard fashion by the inverse of their standard deviation (a combination of the error in the sky and the root-mean-square deviation inside each elliptical annulus).
The errors on the gradients quoted in this paper are the formal errors from the fit.

\subsection{Sky determination and uncertainty}\label{skydetermination}
Arguably the hardest part of the analysis is the determination of the sky background. In our observations, the sky background is made up of zodiacal light and earth shine, but on top of that, diffuse intra-cluster light, tidal debris, scattered light from bright objects and envelopes of big galaxies affect the measured background, so that determining the sky is an extremely difficult problem.

\sex\ provides a measure of the sky and also the standard deviation of the sky. Nevertheless, it is well known that the apertures used by \sex\ are in general too small for large galaxies (see e.g. Paper III), leading to an overestimation of the sky. 

We determine the sky value in the following way. The host galaxy and all other sources in the cutout are masked. We randomly pick $9\times9$ pixels sized boxes in non-masked areas, for which we determine the median. The median of the boxes is then used as the sky value. We find that the uncertainty in the sky is typically 2$\%$ in the F814W (and smaller in F475W), though this number is higher for galaxies near bright sources. We determine the uncertainty in the sky by dividing the width of the histogram by the square root of the number of boxes. We truncate the profiles at the radial distance where the uncertainty in the sky translates into a surface brightness uncertainty of 0.1 mag/arcsec$^2$ (the sky values subtracted by MultiDrizzle are usually of order 25 counts in F814W and 30 counts in F475W, which is 2-3 magnitudes fainter than for ground-based observations of a moonless sky). This means that in a typical case, without bright neighbours, the profiles are truncated at F814W (AB) surface brighness of 25.5 mag/arcsec$^2$ and 26.0 mag/arcsec$^2$ in F475W. \refcom{For a dwarf galaxy this means an extent of about 2 to 3 effective radii, and a bit more for elliptical galaxies.}

\section{Multiple component fits with BAGATELLE}
For all of the galaxies in our sample we determined one-component structural parameters with a Bayesian code called \bagatelle, which fits a parametrized light profile to a galaxy. For the light profile we assumed a S\'ersic profile, where we leave all parameters (S\'ersic index $n$, \Reff, magnitude, ellipticity, position angle, position) free, except the boxiness/disciness parameter, which is fixed to the value that generates perfect ellipses. A more elaborate description of the code can be found in a forthcoming paper (den Brok in prep.), but we summarize some of the characteristics of the code.

Galaxies only rarely consist of one structural component. Even though a S\'ersic profile may be a good approximation for the light profile of the outer galaxy, the centres of galaxies are known to have different profile shapes. For example, dwarf galaxies often show evidence for compact nuclear star clusters \citep{GraGuz03,CotPiaFer06}. Except for the fact that some profiles describe the surface brightness of a galaxy better, several authors have attached physical interpretation to profile shapes as well -- for example, the separation of a galaxy in a S\'ersic profile and a star cluster, or a S\'ersic profile and a core (which may be scoured by black holes.) It is of course important that one uses a profile which describes the data best, however, in the case of 2d profile fitting, the reduced $\chi^2_r$ is usually not a good measure of how well a profile fits. By adding more and more free parameters to a fit, the fit will become better, but probably not more physical.  

This problem can be solved by using Bayesian fitting, because with Bayesian fitting more complicated models (i.e., the ones with more parameters) are punished. We have therefore written a Bayesian code. The input for each model is a set of priors. The output a set of parameters together with a number called {\it evidence}, which tells you how much the prior volume has collapsed. The value of the {\it evidence} can help us choose between models, because it ranks each model in order of likelihood. 

\begin{figure}
\begin{minipage}{82mm}
\center
\scalebox{0.4}[0.4]{\includegraphics{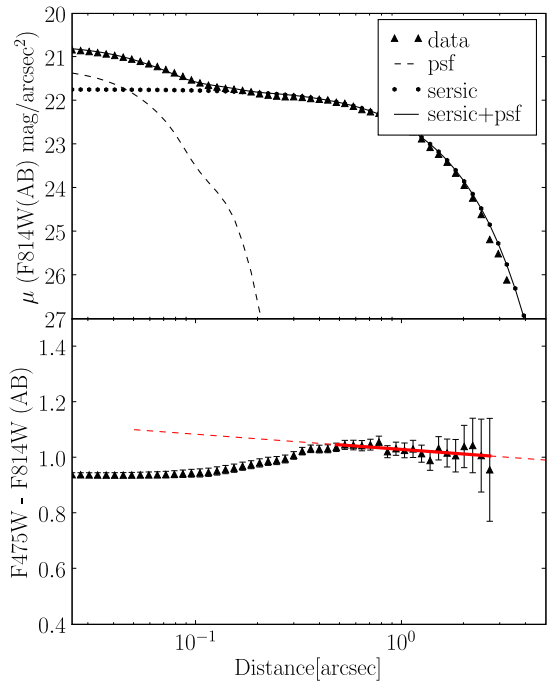}}
\caption{2-Component fit with \bagatelle\ to a Coma dwarf galaxy. Shown are the different components (S\'ersic and point source). The S\'ersic-only fit is not shown, but is according to the difference in {\it evidence} ($\Delta$evidence = 151) extremely unlikely. }\label{fig:bag_prof}
\end{minipage}
\end{figure}

We have fitted our galaxies in Coma with two models. One model consists of a single S\'ersic, the other of a single S\'ersic and a point source, since star clusters at the distance of Coma should not be resolved. The position of the point source is forced to coincide with the position of the rest of the galaxy, so that we do not end up fitting globular clusters. In Figure\ \ref{fig:bag_prof} we show an example fit where the addition of a point source to the profile is a significant improvement on the single S\'ersic fit. We have not fitted our galaxies with additional disc components. For the fainter galaxies in our sample, this is in most cases not a problem (cf. \citet{FerCotJor06}), however, we note that there are S0 galaxies for which the S\'ersic index may be biased. \refcom{We note that the magnitude of NGC4874 is underestimated because of the small field of view of the ACS camera. As this galaxy is a cD galaxy and therefore has a disputable absolute magnitude anyway, we have decided not to correct for this.}

\section{Results}
\subsection{Colour gradients}
In the following sections we present our results on colour gradients. We always use F814W (AB) and F475W (AB) magnitudes. Since B and I$_C$ (in the Johnson-Cousins system) are more commonly used, we give here a few conversions: 
\begin{eqnarray*} 
I_C - \mbox{F814W(AB)} & = & -0.38 
\end{eqnarray*}
(from the WFPC2 Photometry Cookbook)
\begin{eqnarray*}
B & = & \mbox{F475W(AB)} \\ 
& + & 0.329(\mbox{F475W(AB)} - \mbox{F814W(AB)}) + 0.097\\
  & \approx & \mbox{F475W(AB)} + 0.46\\
B - I_C &=& 1.287 (\mbox{F475W(AB)} - \mbox{F814W(AB)}) + 0.538 
\end{eqnarray*}
(from \citet{PriPhiHux09})
\begin{eqnarray*} 
I_C	&=&  I_C(AB) - 0.342\\
B  &=& B(AB) +0.163
\end{eqnarray*}
(from \citet{FreGun94})
\begin{figure*}
\begin{minipage}{135mm}
\center
\scalebox{0.8}[0.8]{\includegraphics{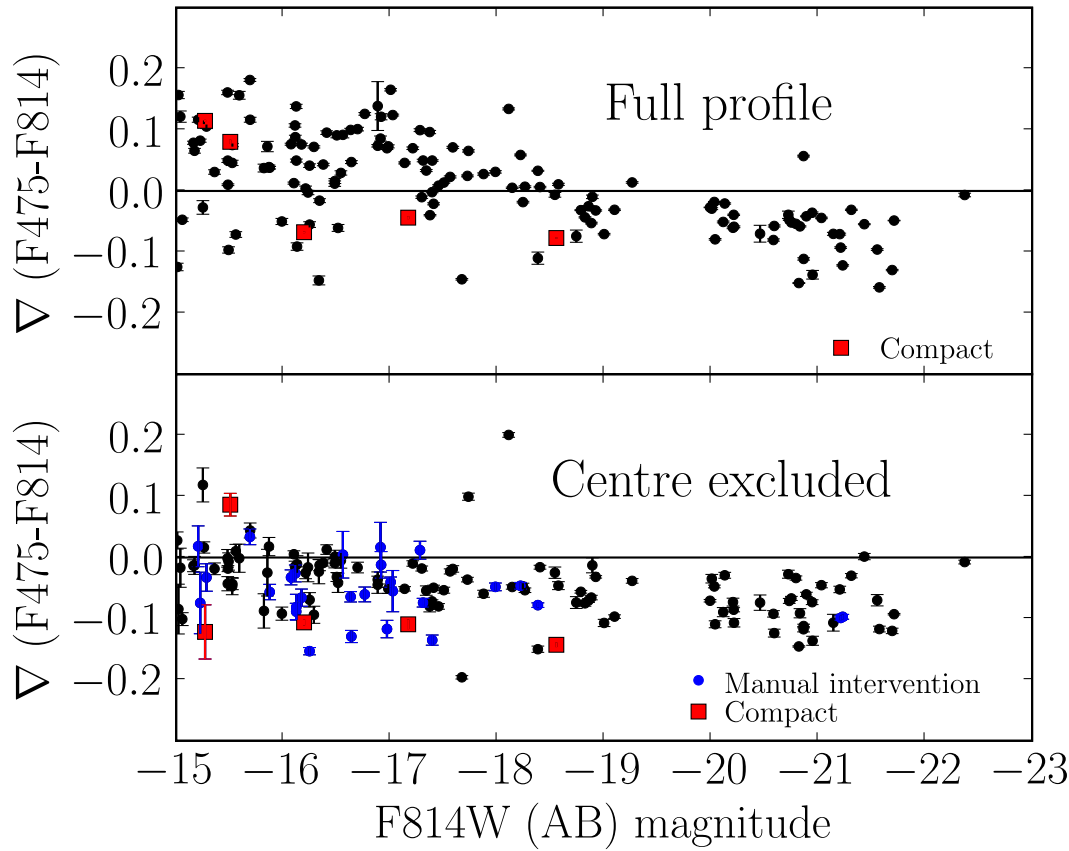}}
\caption{Top: Colour gradients of likely Coma members as a function of absolute magnitude. Bottom: Colour gradients of the same galaxies, now excluding the central parts from the fit. Red squares are compact galaxies from Price et al. (2009). The blue points required manual intervention.}\label{fig:grads_omag}
\end{minipage}
\end{figure*}
In Figure \ref{fig:grads_omag} (top) we show the colour gradient as a function of host galaxy absolute F814W magnitude. Gradients are negative and steep in the large elliptical galaxies, the less luminous galaxies have shallower gradients which are often positive. The change from negative to positive gradients seems to happen just at the magnitude where there is also a well known change in the behaviour of structural parameters of galaxies, where galaxies fainter than $M_V=-18$ (corresponding to $\approx$ $-19$ in F814W) magnitudes follow a different size-luminosity relation than more luminous galaxies (see for example \citet{Kor85}, and for stellar populations \citet{VadVigLac88}). At first glance this looks exactly what one would expect in a galaxy formation paradigm where galaxies form through monolithic collapse.

The centres of galaxies are special environments. Dynamical time scales are short here. Many galaxy centres show evidence for central massive objects such as supermassive black holes and nuclear star clusters. The formation of these objects is not well understood, and in particular their relation with the formation and evolution of the host galaxy is uncertain. If we want to understand what processes have shaped the stellar populations of the host galaxy to its current form, it is necessary to exclude the for this study aggravating contribution of any central AGN or star cluster.

To do so, we carry our linear fits to the colour profile out again, but now excluding a the data within a radius of 3 pixels from the centre (75 pc at a distance of Coma). This is usually enough to exclude most of the light contribution of additional nuclear sources to the colour profile. However, inspection of the colour profile has shown that 3 pixels is sometimes not sufficient ( see for example Fig\ \ref{fig:bag_prof}). For these galaxies, we manually altered the fitting range (maximally excluding 10 pixels from the centre). A more complicated problem is determining gradients in large galaxies with structures such as bars, where an abrupt change in stellar populations between different morphological components can produce a strong colour gradient. These galaxies were also treated by hand, where we always used the outermost component to fit the gradient. 

In Figure \ref{fig:grads_omag} (bottom) we present the colour gradients for galaxies where we have excluded any central component as a function of host galaxy absolute magnitude. This re-analysis has had effect on elliptical galaxies as well as dwarf galaxies. Some elliptical galaxies have now less steep colour gradients, because we eliminated central reddening due to dust discs, whereas the to a large extent positive gradients that we observed for dwarf galaxies in Figure~\ref{fig:grads_omag} (above) have mostly disappeared and become slightly negative. There are two galaxies with strong positive gradients. One of these galaxies (\texttt{COMAi13005.684p275535.20}) shows a central disk and spiral structure. This galaxy may be a spiral galaxy transforming to a dwarf galaxy, even though its projected distance is not too far away from the core of the cluster.

Even without the positive gradients for dwarf galaxies, the trend is visible that fainter galaxies have smaller gradients (a Spearman rank test confirms this with a correlation coefficient of 0.41 (excluding the compacts gives 0.45)). This is at least qualitatively consistent with a monolithic collapse scenario. In this scenario, after a first burst of star formation, primordial gas is enriched with stellar ejecta until the onset of galactic winds prohibits any further star formation. Since the latter process is dependent on the local escape velocity, colour gradients are a function of host galaxy mass, for which we use absolute magnitude as a proxy in Figure \ref{fig:grads_omag}.

In a subsequent paper, we will investigate the relationship between the colour of the nuclear region with that of the main body of the galaxy. In the following part of this paper, we focus only on the gradient in the main body of the galaxy, i.e. excluding the central part.

\subsection{Differences in colour gradients as function of environment}
\begin{figure}
\begin{minipage}{82mm}
\center
\scalebox{0.4}[0.4]{\includegraphics{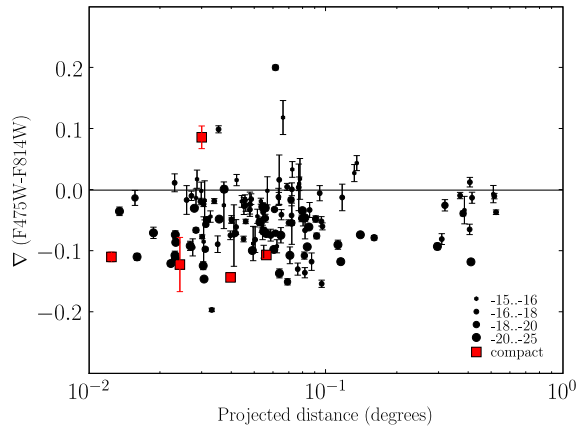}}
\caption{Colour gradients of likely Coma members as function of harmonic mean distance towards NGC4874 and NGC4889. Red squares are the compact galaxies identified by Price et al. The size of the dot is related to the magnitude of the host galaxy as indicated in the figure.}\label{fig:grad_dist}
\end{minipage}
\end{figure}

\begin{figure}
\begin{minipage}{82mm}
\center
\scalebox{0.4}[0.4]{\includegraphics{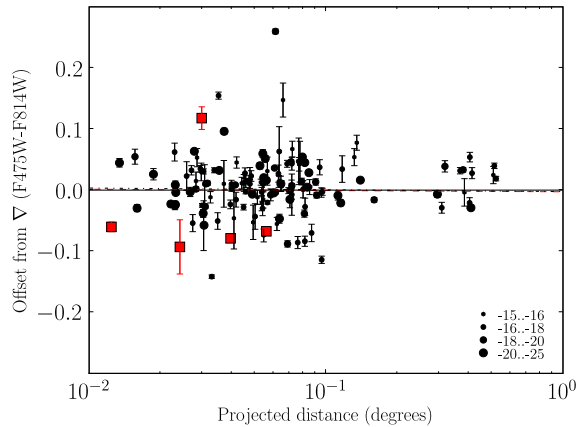}}
\caption{\refcom{Offset of the colour gradients from a linear fit to the data in Fig.~\ref{fig:grads_omag} (bottom panel) plotted against the harmonic mean distance towards NGC4874 and NGC4889.} Compact galaxies are red. The lines are fits through the sample (the red dashed line includes the compacts, the almost indistinguishable black dashed-dotted line shows the fit after including the compacts.)}\label{fig:grad_dist_corr}
\end{minipage}
\end{figure}

The environment in which a galaxy evolves can influence its colour gradient, since it can regulate the amount of star formation (by starving a galaxy from an external gas reservoir, by removing already present gas, by heating or ionizing cold gas or by compressing present gas due to the high-pressure in the ICM) and set the number of interactions with other galaxies, since harassment and mergers can alter the morphological appearance of a galaxy.
One therefore expects to see differences between galaxies in the outskirts of the cluster and galaxies near the centre of the cluster.
In Figure~\ref{fig:grad_dist} we plot the colour gradients as a function of clustercentric distance. Although the results are not particularly sensitive to the way the distances are combined, we adopt as our indicator the harmonic mean of the distances to the two brightest galaxies, NGC4874 and NGC4889.
It is difficult to find any environmental trend in Figure~\ref{fig:grad_dist}. Indeed the value of Spearman's rank correlation coefficient is 0.08 (0.06 when excluding the compacts) which means that there is no correlation at all.
It is known from our previous analysis that more massive galaxies have stronger colour gradients. On the other hand, the clustering properties of massive galaxies are different from those of dwarf galaxies, in the sense that low-mass gas-poor galaxies are generally more strongly clustered than high-mass gas-poor galaxies. A better question to ask is perhaps if the colour gradients of galaxies with similar mass are somehow dependent on density.
We therefore decided to divide out the gradient-magnitude relation for bright galaxies, by fitting a line to the data of Figure~\ref{fig:grads_omag}. \refcom{The residuals of this fit, i.e. the vertical offsets of the colour gradients from this line, are shown in Figure \ref{fig:grad_dist_corr} as a function of distance}. \refcom{To check if there is any environmental dependence on the offsets, we have fitted a line through them, as a function of logarithmic projected distance. We find a slope consistent with zero, meaning that there is no evidence in our data that the offsets from the main relation are due to environment.} (Even when excluding the compact galaxies from the fit, the reduced $\chi^2$ remains high and the slope is consistent with zero). A Spearman correlation coefficient test shows no correlation ($r=0.06$ or $0.04$ without compacts.)  

\section{Comparison with structural properties}
\label{chap:struc}
The sequence of giant elliptical galaxies can be divided into two classes: the rapidly rotating low-luminosity ellipticals which are quite flattened, and their (generally) higher luminosity slow- or non-rotating counterparts. Despite their dynamical and morphological differences, elliptical galaxies are fitted exceptionally well by S\'ersics $R^{\frac{1}{n}}$ surface brightness law, where $n$, known as the S\'ersic index correlates positively with galaxy magnitude and effective radius.

\subsection{Correlation with structural parameters}
\begin{figure*}
\begin{minipage}{135mm}
\center
\scalebox{0.8}[0.8]{\includegraphics{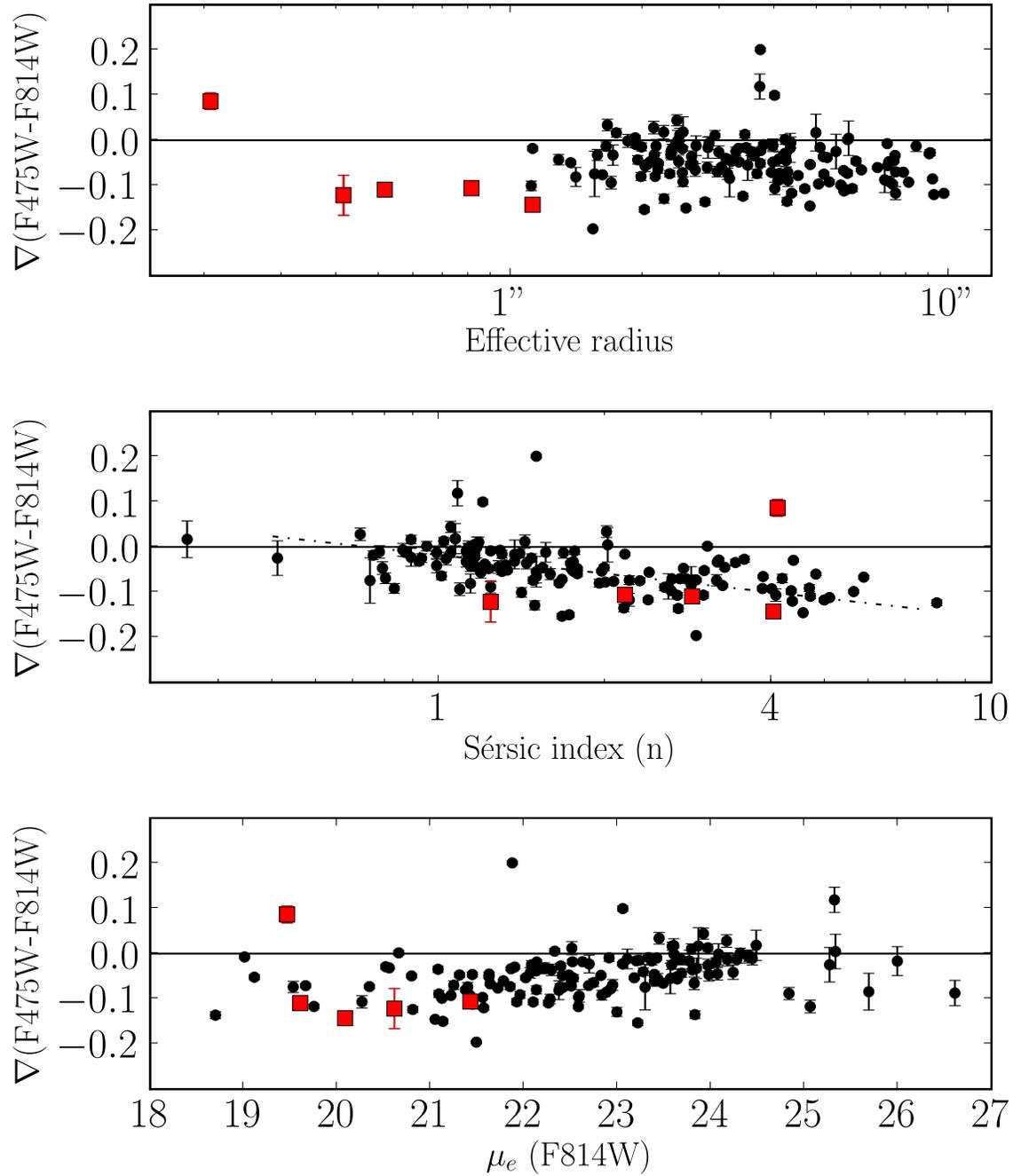}}
\caption{Colour gradients of likely Coma members, plotted against effective radius, S\'ersic index {\it n} and surface brightness at the effective radius. All quantities are calculated in the F814W band. The dashed-dotted line ($\nabla = -0.018 -0.1372\log{n}$) in the middle panel is a fit to the data, excluding gradients steeper than $+0.15$ and compact galaxies.}\label{fig:grads_struc}
\end{minipage}
\end{figure*}

In Figure \ref{fig:grads_struc} we show how the colour gradients correlate with the structural properties that we have derived with our 2d Bayesian fitting code. We have used the results of the S\'ersic component from the best fitting model. If one only fits single-S\'ersic components to galaxies, one often finds a too high S\'ersic index for dwarf galaxies.
A lot of the behaviour in this plot is expected. Smaller galaxies have less steep colour gradients (this was expected because fainter galaxies are generally smaller). Similar behaviour is observed in the panels where we plot the gradient against S\'ersic index and effective surface brightness.
We have refitted the galaxies that were identified as compact ellipticals by \citet{PriPhiHux09}. The offset in the compact sources is highest in the effective radius panel. Surprisingly, the one compact source with a positive gradient appears roughly consistent with the other sources.

\section{Comparison with literature data}
The colour of a stellar population is a function of both age and metallicity. Age, metallicity as well as other line index gradients have been studied. In this section we compare our colour gradients with Mgb gradients and metallicity gradients, derived by various authors in the literature. The advantage of comparing directly with index gradients is that we compare direct observables with each other. \refcom{Comparison with metallicity gradients involves a model to convert colour to metallicity, and requires assumptions on age gradients. For ages greater than 2 Gyr, and metallicities higher than $[\mbox{Fe}/\mbox{H}]=-0.6$, which is definitely the case for our comparison galaxies, the conversion between colour and Mgb is completely linear, and independent of age and metallicity.} 
\begin{figure}
\begin{minipage}{82mm}
\center
\scalebox{0.4}[0.4]{\includegraphics{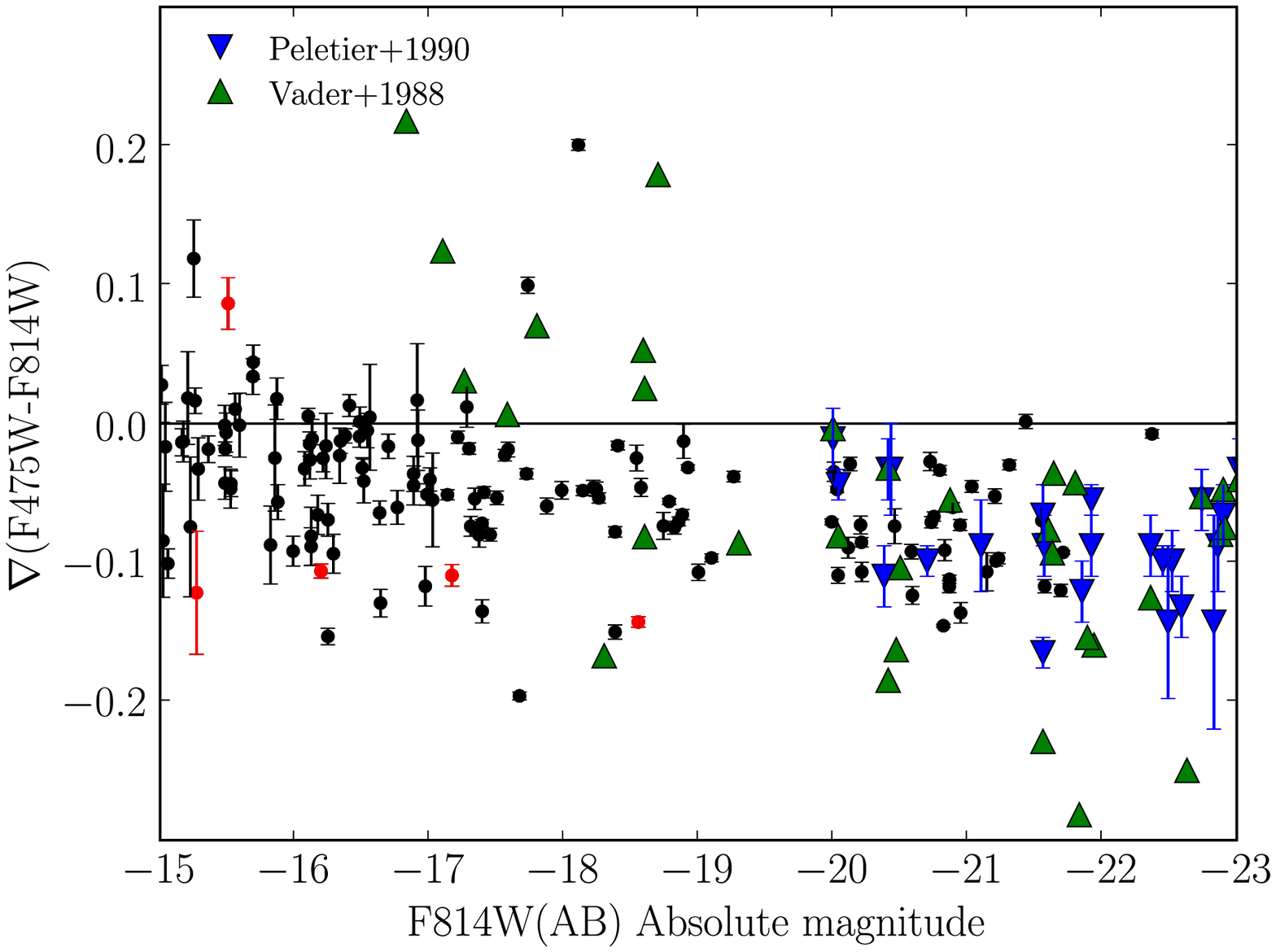}}
\caption{Comparison of our colour gradients in the outer parts with the work of \citet{VadVigLac88} and \citet{PelDavIll90}.}\label{fig:comparison_peletier}
\end{minipage}
\end{figure}

However, for the interpretation of our colour gradients we also compare them to metallicity gradients. \refcom{A comparison with previous work on colour gradients is shown in Fig.~\ref{fig:comparison_peletier}. The data shown there are from \citet{VadVigLac88} and \citet{PelDavIll90}. The B-band magnitudes are taken from \citet{deVdeVCor91} and converted to absolute magnitudes using the distances from the NASA/IPAC Extragalactic Database (NED). There is clearly a big improvement in the range of magnitudes that is covered by our data. We do not confirm the positive colour gradients for dwarf galaxies as found by \citet{VadVigLac88}, but our data are consistent with those of \citet{PelDavIll90}. It is likely that previously correlations were not discovered because of this smaller magnitude range. However, it has been found by other authors that there is a turnover in population gradients around $M_{F814W} \approx -22.$ \citep[for example, ][and references therein]{SpoProFor09}. Unfortunately our our sample lacks galaxies in this magnitude regime.}
Since the galaxies with which we compare are quite large, seeing effects for the spectroscopic data should be negligible and exclusion of the central parts should in most cases not be necessary, as most of the bright galaxies do not have nuclear star clusters. 
For most of these data, only Johnson B magnitudes are available. We convert these to F814W magnitudes using $\mbox{B}-\mbox{F814W}=1.8$ This is not entirely accurate, but the error which we make with this conversion should be less than 0.2 mag for quiescent galaxies which are not affected by dust or recent star formation, and since we care mostly about global trends, we find this error acceptable.  
\begin{figure*}
\begin{minipage}{185mm}
\scalebox{0.45}[0.45]{\includegraphics{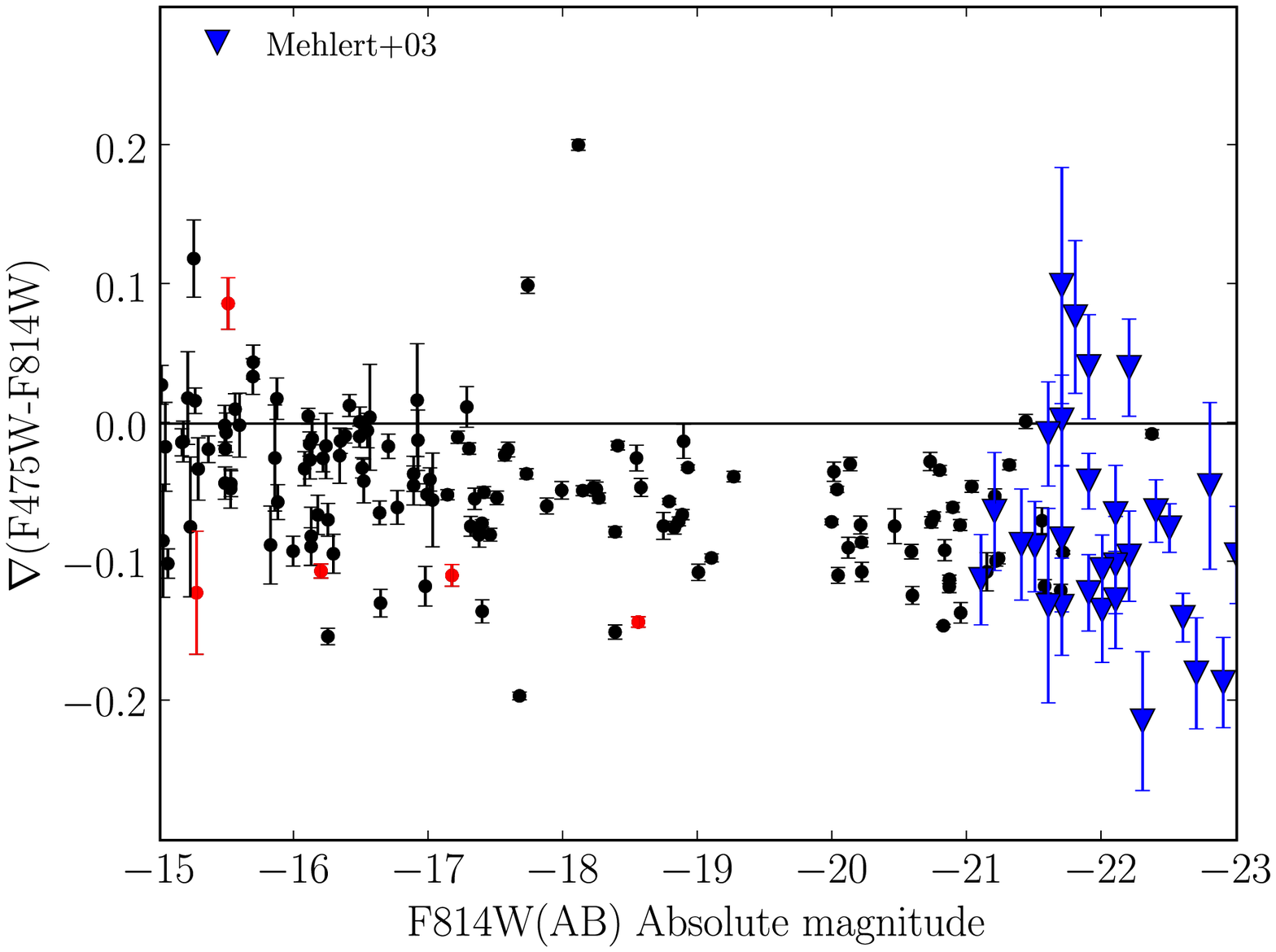}}\scalebox{0.45}[0.45]{\includegraphics{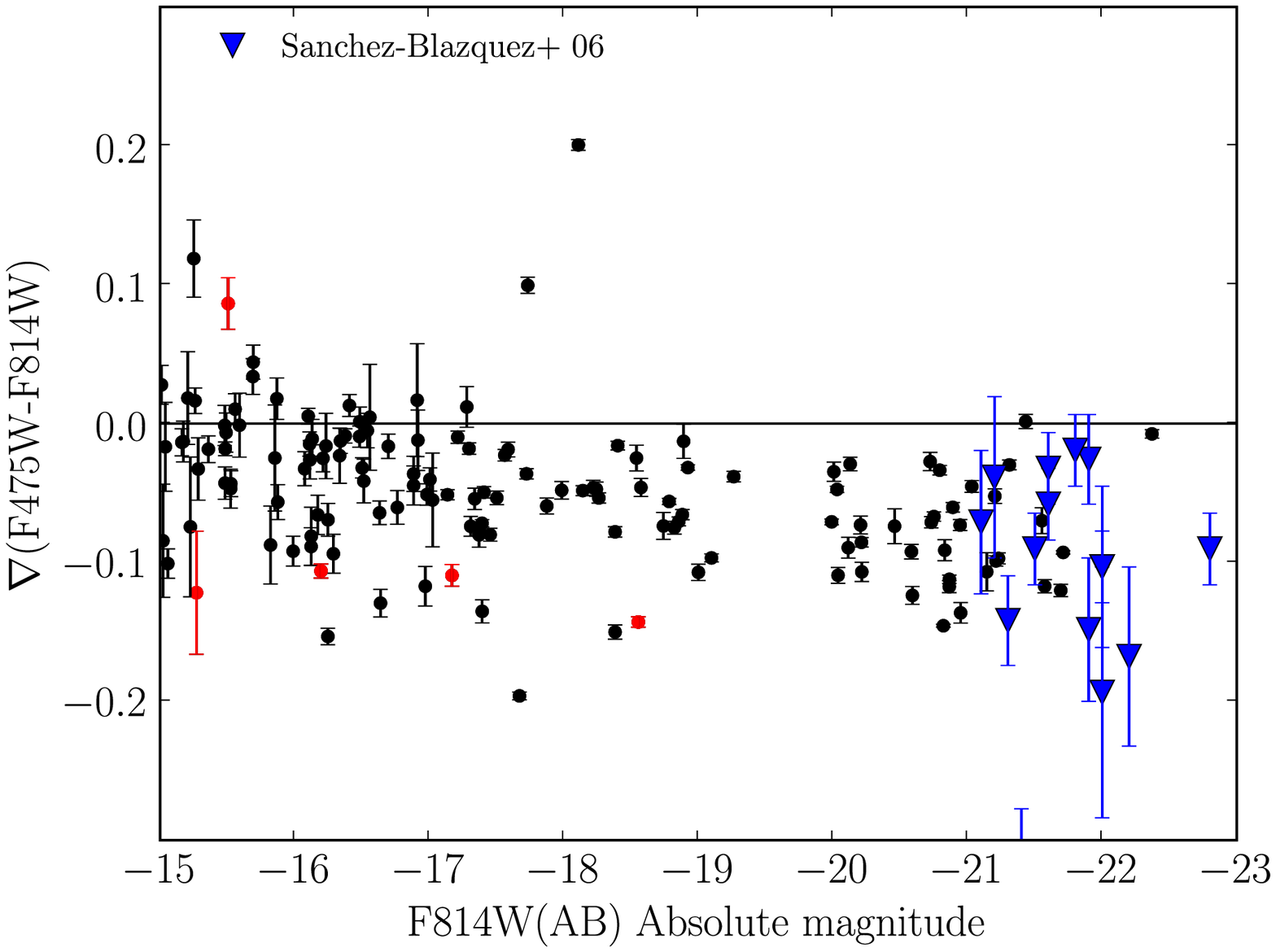}}
\scalebox{0.45}[0.45]{\includegraphics{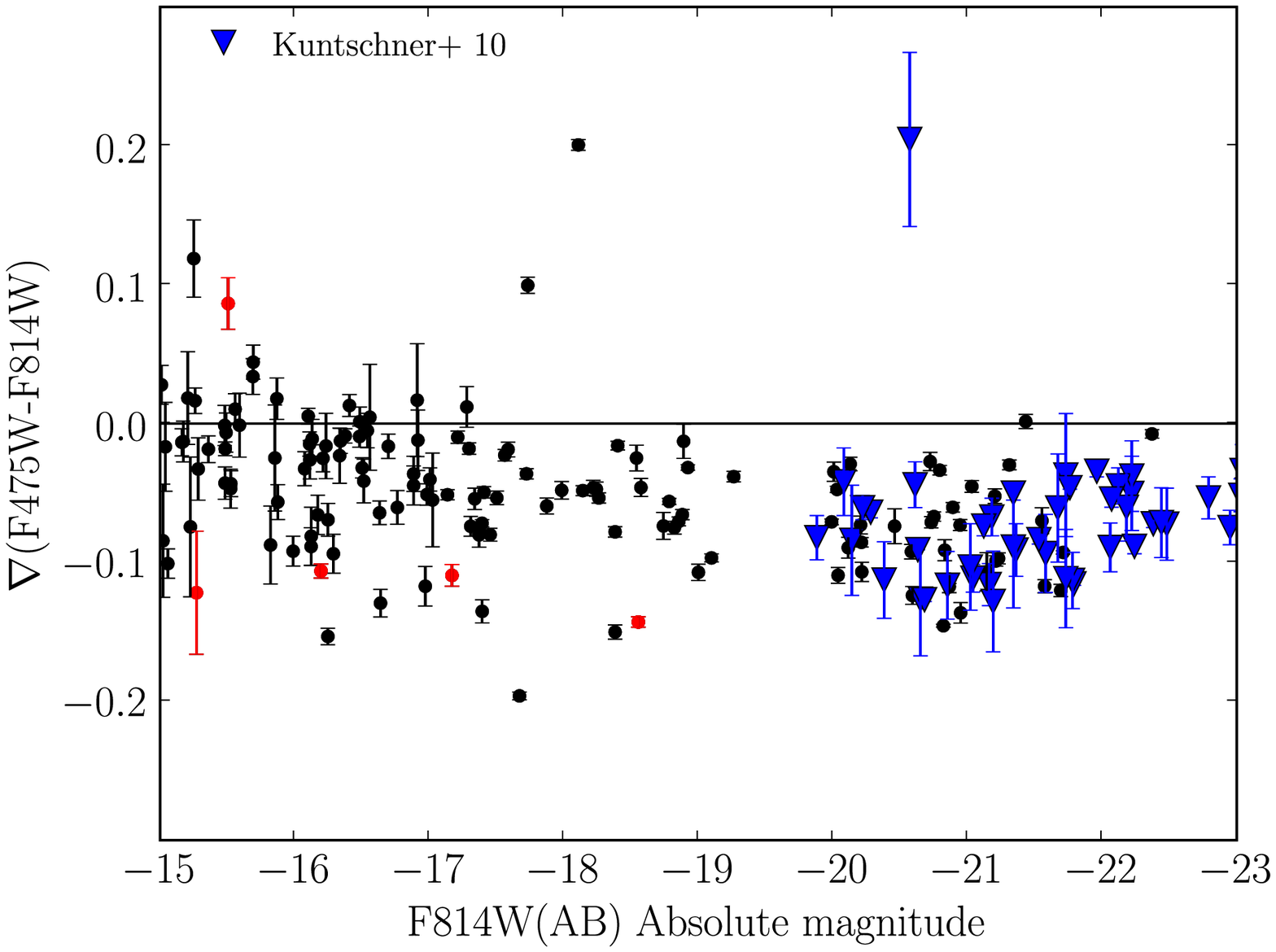}}\scalebox{0.45}[0.45]{\includegraphics{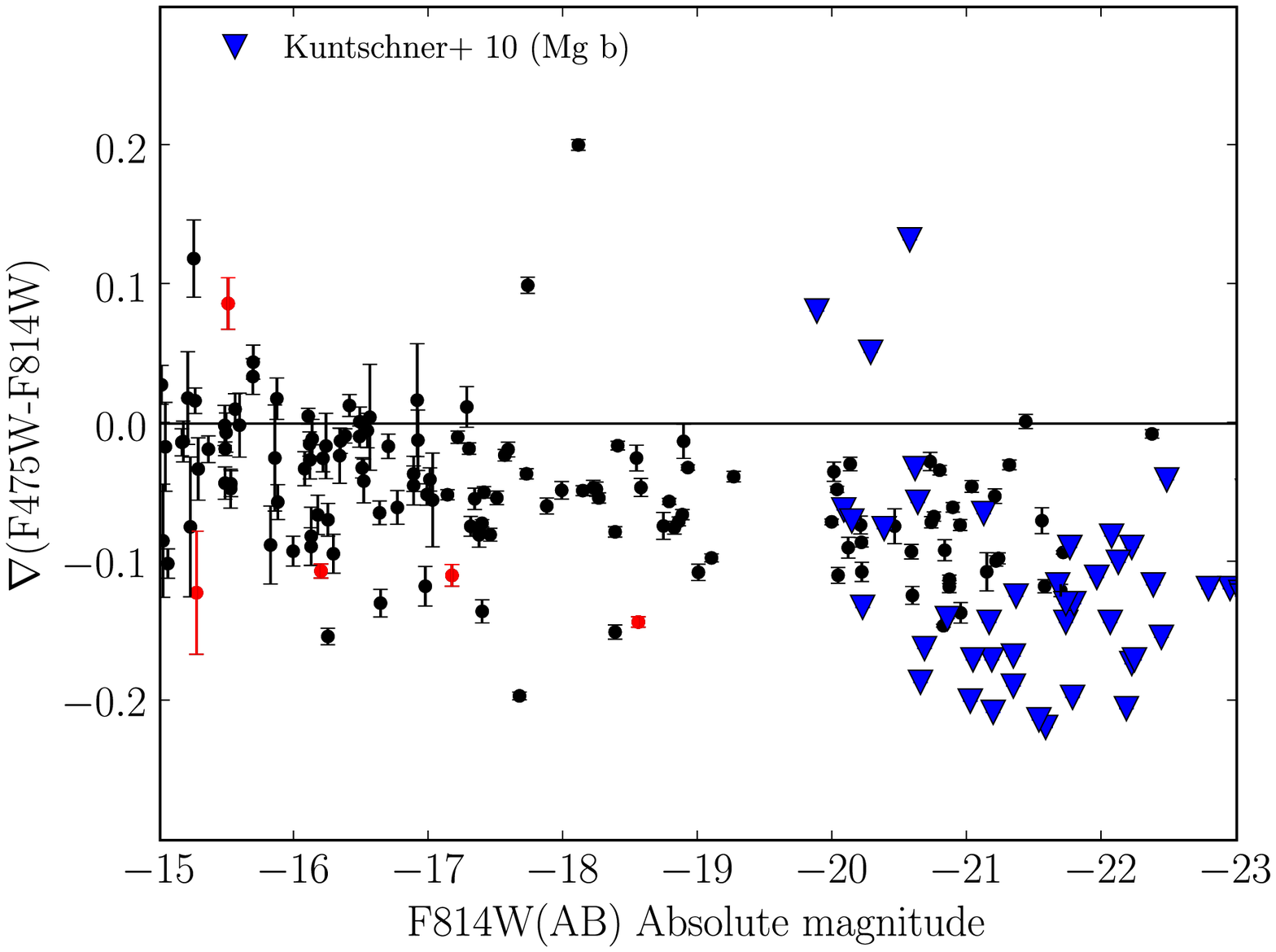}}
\caption{Colour gradients of likely coma members (black points) compared to metallicity and Mgb gradients from the literature (blue triangles). Upper panels: (left) Mgb data from Mehlert et al. (right) Mgb data from Sanchez-Blazquez et al. Lower panels: (left) Metallicity data from Sauron (Kuntschner et al. 2010) (right) Mgb data from Sauron (Kuntschner et al.) The Mehlert and Sanchez-Blazquez datasets plotted here consist entirely of Coma galaxies. However, the Sauron data has [Z/H] gradients compare better with our observations.}\label{fig:literature_comparison}
\end{minipage}
\end{figure*}
\subsection{Comparison with Mehlert et al. 2003}
\citet{MehThoSag03} fit line index gradients to data obtained by long-slit observations for a sample of 35 early-type galaxies in Coma. A benefit of comparing our data with this sample is, that we are sampling objects from a similar environment, and any systematic error in the distance is the same for both samples.    
We used the {\texttt galaxev} software \citep[henceforth BC03]{BruCha03} to convert colours to an Mgb index.  We measure the Mgb indices with the same index definition in {\texttt galaxev}, using the Padova 1994 evolutionary tracks and a Salpeter IMF (lower and upper mass cut-offs at 0.1 and 100 Msun). From this analysis we find:
\begin{eqnarray}
\frac{\Delta \mbox{Mgb}(\mbox{\AA})}{\Delta \log(r)} = 0.244\frac{\Delta \mbox{F475W}-\mbox{F814W}}{\Delta \log(r)}
\end{eqnarray}
Mehlert et al. define gradients in indices however as:
\begin{eqnarray}
\Delta(Index) = \frac{\Delta\log{Index}}{\Delta\log{R}} = \frac{\ln(10)}{Index} \frac{\Delta Index}{\Delta\log{R}}
\end{eqnarray}
Which means that there is an additional metallicity factor in the conversion from colour gradient to metallicity gradient, for which we use the signal-to-noise averaged central value from the same paper. \refcom{A different way of doing this is fitting a relation between $\log(\mbox{Mg} \mbox{b})$ and colour, but this gives essentially the same answer.} Figure~\ref{fig:literature_comparison} shows a comparison between our derived gradients and the Mgb-derived colour gradients. The comparison shows that the scatter in these data is higher than in our data. Moreover, some gradients are positive, whereas the colour gradients in this magnitude regime are exclusively negative. 

\subsection{Comparison with Sanchez-Blazquez et al. (2006)}
\citet{SanGorCar06} made an extensive study of metallicity gradients in early type galaxies, both for galaxies in dense regions and field galaxies. Their sample includes also 21 members of the Coma cluster. Even though their full sample is much larger, again, to sustain homogeneity of the sample and avoid any additional uncertainties in absolute magnitudes due to distance uncertainties, we focus exclusively on their 21 Coma cluster galaxies.
Because the step from index gradients to metallicity gradients requires a lot of assumptions on stellar models, we again just use the Mgb gradients measured by these authors. Because the gradient and index definition of Sanchez-Blazquez et al. is slightly different from Mehlert et al., the conversion that we use here is just the simple:
\begin{eqnarray}
\frac{\Delta \mbox{Mgb}(\mbox{mag})}{\Delta \log(r)} = 0.155\frac{\Delta \mbox{F475W}-\mbox{F814W}}{\Delta \log(r)}
\end{eqnarray}
although in fact we use the inverse relation (i.e. we predict a colour from a given metallicity gradient). 
In the upper panel of Figure\ \ref{fig:literature_comparison} we compare the predicted colour gradients from the Mgb line strengths to our observed colour gradients. The converted gradients from Sanchez-Blazquez et al. are consistent with our observed gradients, except for two galaxies, which would have colour gradients steeper than $-0.25$ mag/dex.  

\subsection{Comparison with SAURON}
\citet{KunEmsBac10} give a number of different line strength indices for early type galaxies in the SAURON sample, from which they determine metallicity and age gradients. Their data are obtained using integral field spectroscopy, and have very high signal to noise. The SAURON sample consists of a magnitude limited sample of 48 early-type galaxies in the northern hemispere with $v_{\mbox{hel}} < 3000\mbox{km/s}$. 

Their fits are carried out in the same way as ours, by fitting a logarithmic gradient to the 2-dimensional metallicity maps, which were determined by fitting SSP models to different spatial bins.
The SAURON sample contains galaxies from nearby clusters and from the field, so a different density regime is probed by these observations. All galaxies from the sample are well-studied nearby galaxies, so we assume that the distance uncertainty is small.

Figure \ref{fig:literature_comparison} (above) shows a comparison of the SAURON observations with our colour gradients. 
Amazingly, the data follows the same trends as ours: it shows a trend for gradients to become flatter for galaxies fainter than absolute F814W magnitude fainter than $-20$, the very shallow gradient for the very brightest galaxies (although we only have one data point) and also the scatter and magnitude of the gradients are completely consistent (if we exclude the one point with a very strong positive gradient.) 

We can also compare our colour gradients with the Mgb data from the same paper. Mgb is measured at two locations in each galaxy, at R$_e$ and R$_e/8$. The fact that these gradients are based on only two points, one of which is measured relatively close to the galaxy center, makes these gradients extremely sensitive to the presence of central stellar populations which have formed much later than the galaxy halo, and hence the galaxies can show very strong gradients. In addition to this, Mgb gradients can also become stronger if the $\alpha/Fe$ ratio increases toward the galaxy centers. 
The results in Figure \ref{fig:literature_comparison} show that the Sauron data have a low intrinsic scatter, which may be due to higher signal-to-noise than for example the Sanchez-Blazquez data. Also, the Sauron [Z/H] and Mgb gradients do not show the same trend. \refcom{We conclude that at least in this magnitude regime our interpretation of colour gradients as metallicity gradients seems to be consistent with spectroscopic data.}

\subsection{Comparison with Spolaor et al.}
\begin{figure*}
\begin{minipage}{135mm}
\center
\scalebox{0.8}[0.8]{\includegraphics{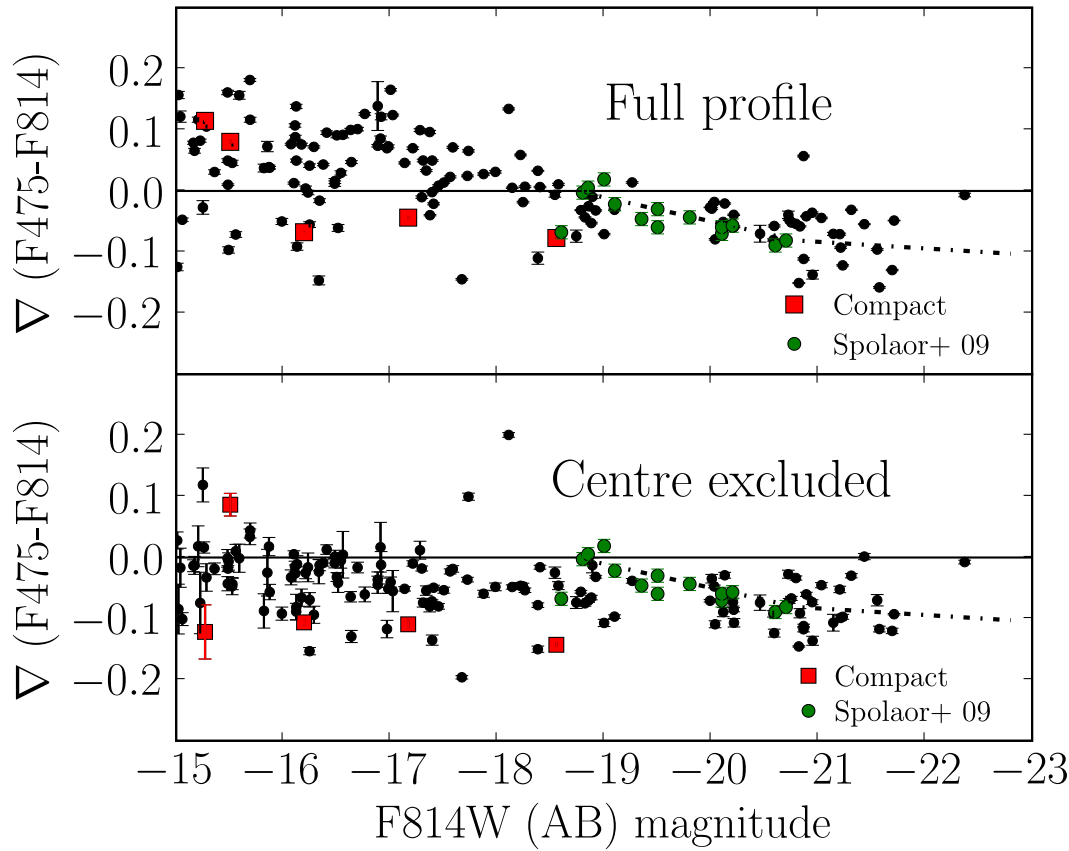}}
\caption{Top: Colour gradients of likely Coma members and converted Spolaor et al. (2010) data. Bottom: Colour gradients after excluding the central part as function of absolute F814W magnitude. The blue points are the points from Spolaor et al., the dashed line is the theoretical prediction from Kawata \& Gibson (2003)}\label{fig:spolaor}
\end{minipage}
\end{figure*}

\citet{SpoProFor09} use GMOS in long-slit mode to measure spatially resolved line indices in 6 Fornax and 8 Virgo cluster galaxies. They convert their Lick/IDS indices to metallicies and ages using the SSP models from \citet{ThoMarBen03,ThoMarKor04}.  
The metallicity gradients that they find drop from [Z/H]=$-0.3$ to 0 for fainter galaxies. The velocity dispersion-gradient relation found by these authors shows a surprisingly small scatter. We convert the metallicity gradient -- B-magnitude data using BC03 to colour gradient -- I-magnitude data. For this we use B(J) $-$ I(AB)=1.8 and $\Delta$F475W $-$ F814W = 0.27 $\Delta$[Z/H] The results of this conversion are shown in Figure\ \ref{fig:spolaor}, where we also overplot the theoretical predictions from \citet{KawGib03}.   
Surprisingly, we find, if we compare the gradients with the results from Spolaor et al., that the results are consistent only if we use the full gradient, and not excluding the centre.

\subsection{Comparison with Koleva et al.}
The comparison data in the previous subsections was mainly focussed on bright ellipticals. However, an important result of our work is that dwarf galaxies have slightly negative gradients. In this section we make a detailed comparison between our gradients and the data from \citet{KoldeRPru09}, a paper studying metallicity gradients in dwarf galaxies. These authors obtained FORS1/2 long slit observations of a number of galaxies in the Fornax cluster, the Antlia cluster and the NGC5044 and NGC5898 groups. After rebinning to obtain sufficient S/N in the outer parts, they  fit SSP models as a function of galactic radius to obtain ages and metallicities.

\begin{figure}
\begin{minipage}{82mm}
\center
\scalebox{0.4}[0.4]{\includegraphics{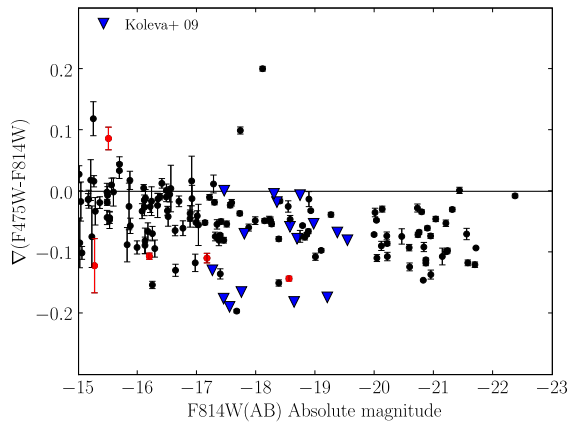}}
\caption{FORS metallicity gradients determined from data from Koleva et al. converted to colour gradients. In black we show the data from the ACS Coma Cluster survey.}\label{fig:comparison_koleva1}
\end{minipage}
\end{figure}

Inspection of the data reveals that most age gradients are negligible, except for the centre -- where dwarf and low mass elliptical galaxies usually host a star cluster -- which is often, but not always, younger. We fit a gradient to the metallicity data, which we subsequently convert to a colour gradient using the BC03 models.
Figure~\ref{fig:comparison_koleva1} shows the gradients of the Fornax data compared to our own data. Remarkably, part of the galaxies seems to be consistent with our own gradients, whereas a group of points seems to be more consistent with the compact sources in Coma. 

\begin{figure}
\begin{minipage}{82mm}
\center
\scalebox{0.4}[0.4]{\includegraphics{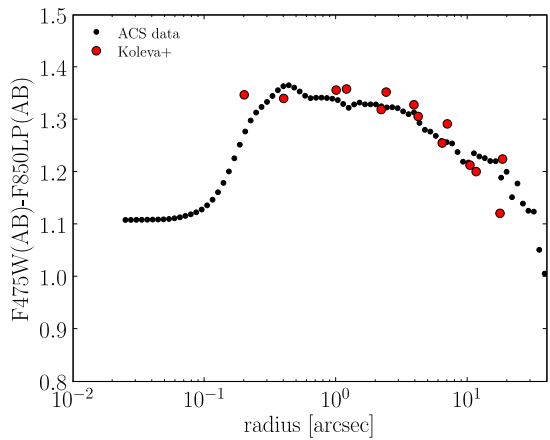}}
\caption{FORS metallicity data from Koleva et al. converted to colour data (red) with data from the ACS FORNAX survey in black.}\label{fig:comparison_koleva2}
\end{minipage}
\end{figure}

For one Fornax galaxy, FCC136, ACS data is available from the Fornax Cluster ACS Survey \citep{JorBlaCot07}, in the F850LP and F475W passbands. The wavelength of the F850LP is a bit redder than our own ACS data, but this offset should be negligible for gradients. 
In Figure~\ref{fig:comparison_koleva2} we show a detailed comparison between the metallicity points and colour of FCC136. According to the data from Koleva, this galaxy has no age gradient, and put even more strongly, should even be a true SSP. 
Inspecting the colour profile shows that there is a nuclear star cluster present, which is much bluer than the rest of the galaxy, either because of a lower metallicity, or because of a younger age. It is likely that the centre contains a much younger population that could not have been seen from the ground.
 
Outside the centre, the metallicity profile does seem to follow the colour profile, with a small offset, which may be caused by uncertainties in the stellar population models. Foreground extinction in the direction of Fornax is small (around E(B$-$V)=0.0155 according to \citet{SchFinDav98}) and hence E(g$-$z) should be around 0.033 \citep{SirJeeBen05}. This is a constant, but somewhat uncertain colour offset, and we do not correct for it. (The small wiggle and subsequent flattening in the profile are due to use of a slightly too small PSF in the red band.) This plot shows also that one should be very careful in deciding which points to take into account for the fit.

\section{Discussion}
\subsection{Metallicity gradients in dwarf galaxies}
One of the shortcomings of using broad-band data for studying gradients in galaxies is that one cannot separate age and metallicity contributions to the gradient. Both mechanisms are known to operate in galaxies. For example, M32 has flat colour gradients in most colours \citep{Pel93,LauFabAjh98} but shows both an age and metallicity gradient in long slit observations in the sense that the galaxy is older and more metal poor on the outside \citep{RosAriCal05}. However, M32 is classified as a compact elliptical and is probably not representitive for the class of red-sequence galaxies as a whole. 
Is there any evidence for age gradients in 'normal' red-sequence galaxies?  \citet{KunEmsBac10} determine age gradients for galaxies in the Sauron sample and find that old (SSP age$>$8Gyr) galaxies are consistent with having no age gradients, but young galaxies often have a younger centre. Sanchez-Blazquez et al. do find age gradients for their sample of galaxies in low density environments, but their results for galaxies in a high density environment (Coma) point at age gradients consistent with zero. This means that at least for the bright galaxies in our sample, the age gradients are probably (on average) not important. However, we note that {\citet{RawSmiLuc10} warns that, even though they find on average zero age gradients, 40\% of their sample is inconsistent with the absence of an age gradient.\\ 
The situation may be very different in dwarf galaxies. \citet{StiDalQui09} simulate age gradients in isolated dwarf spheroidal galaxies (dSphs) and find haloes which are systematically older than the inner parts of the galaxy. Also other hydrodynamical simulations of dwarf ellipticals \citep[see e.g.][]{ValdeRDej08} show that star formation proceeds in bursts of gas that contracts towards the galaxy centre, with each subsequent starburst being more central, and even within one burst small age gradients are produced. The age gradient that is formed in this way is positive, and would lead therefore to a positive colour gradient. There has not been much observational work to test this with dwarf ellipticals. \citet{KoldeRPru09} measure gradients in bright dEs in the Fornax cluster, although they claim that most galaxies have positive age gradients, excluding the centre from the analysis the profiles look remarkably flat. Also the results of \citet{SpoKobFor10} do not point at any age gradients. On the other hand, results for dSphs in the local group (see e.g. \citet{BatTolHel06}) and the M81 group \citep{LiaGreKoc10} point at slightly older outer haloes of galaxies.

If we assume that dwarf galaxies form outside-in (and this is where all the evidence from simulations points to), the fact that we see primarily negative colour gradients in dwarf galaxies tells us the following two things: dwarf galaxies do have metallicity gradients and the contribution of the age gradient to the colour gradient is, although maybe not negligible, not as strong as the metallicity gradient. On the other hand, one of the two galaxies that has a strong positive gradient may be a spiral galaxy transforming to a dwarf elliptical. The inverted gradient here may be due to recent star formation, so age gradients may be present if dwarfs are stripped spiral galaxies. 
  
\subsection{Comparison with theoretical models}
In the previous sections we have derived colour gradients for a homogeneous sample of Coma dwarf and elliptical galaxies. We have shown that the strength of the gradient is strongly dependent on the measurement method: including the centre alters the gradients significantly. If we however exclude the centre, motivated by the fact that galaxy centres host cores, light excess, dust and star clusters, which, although being consequenses of evolution, are probably not generated by the initial collapse of the galaxy (assuming that a galaxy forms through a monolithic dissipational burst), the colour gradients stay slightly negative.
The reduced gradients show a relation in which gradients become less steep towards faint magnitudes, but almost never become positive. This is at least qualitatively consistent with the old monolithic collapse scenario. Here, an elliptical galaxy forms as a large starburst from a primordial gas cloud. As not all the gas is used in the initial burst, the remaining gas sinks to the centre, whereas the stars stay where they are. On its way to the centre, the gas is enriched by stellar ejecta. In this way, a metallicity gradient is formed, with the stars near the galactic centre becoming more metal rich. If any age gradient is present, it should be a positive one, but the metallicity gradient set up by the collapse is usually much stronger in the colour, and there is no strong observational evidence for age gradients. \citet{Car84} was one of the first to model dissipational collapse of an elliptical galaxy with an N-body code. In his simulation collisional gas cloud particles collide with each other to form collisionless star particles. The metallicity gradient of the simulated galaxies depends chiefly on the assumed ratio of the velocity of metal-enriched stellar ejecta and the velocity dispersion of the stars and has a value of ${\nabla}\mbox{Z}=-0.5$ for typical elliptical galaxies. We note however that lowering this velocity dispersion yields gradients that are more consistent with our observations, and we indeed confirm a trend that fainter galaxies have less negative gradients.  

Since then models have improved and are better at reproducing observations. \citet{ChiCar02} carried out N-body-tree-SPH simulations, thereby aiming at reproducing the properties of dwarf galaxies. Their simulations start from a virialized proto-galaxy with a $1/r$ dark matter density profile. One of their conclusions is that galactic winds are extremely important in blowing out metal rich gas from a galaxy and galactic winds of low mass galaxies are more efficient. From their Figure 6 we reconstruct metallicity gradients of approximately $-0.8$ for a $10^{12}$ M$_\odot$ galaxy and essentially zero for a $10^{9}$ M$_\odot$ galaxy.

There have been many more attempts to model metallicity gradients, also in a cosmological context. The simulations by \citet{KawGib03} model elliptical galaxies by following an isolated sphere with cold dark matter density fluctuations superimposed. These simulations show the same trend from Chiosi \& Carraro that less massive galaxies have less steep gradients, which even go to zero for the least massive galaxies. Their simulation output for metallicity gradients does however contain only three different mass points for elliptical galaxies. Most recently, \citet{PipDErMat08,PipDErChi10} modelled metallicity gradients in elliptical galaxies. These authors reached the conclusion that the simple outside-in formation of ellipticals is insufficient to explain observables and requires more physics, such as detailed modelling of gas flows. In stark contrast with the models from \citet{KawGib03}, the Pipino et al. models show a variety of gradients, strongly dependent on the star formation efficiency. The galaxies with high SF efficiency have steeper gradients.
Figure~\ref{fig:pipino} compares our observed gradients with the theoretical models by Pipino et al. Their paper does not provide magnitudes of galaxies, so we make a simple conversion to F814W magnitudes using the Miles\footnote{\texttt{http://miles.iac.es}} SSP models \citep{VazSanFal10} where we assume a single age (10 Gyr) and metallicity ([Fe/H]=0) for all galaxies. Since we are only interested in the qualitative properties, we do not correct for the fact that smaller galaxies have on average a lower metallicity. The range of metallicity gradients predicted by these models seems to coincide well with our observations. 

\begin{figure*}
\begin{minipage}{162mm}
\center
\scalebox{0.72}[0.72]{\includegraphics{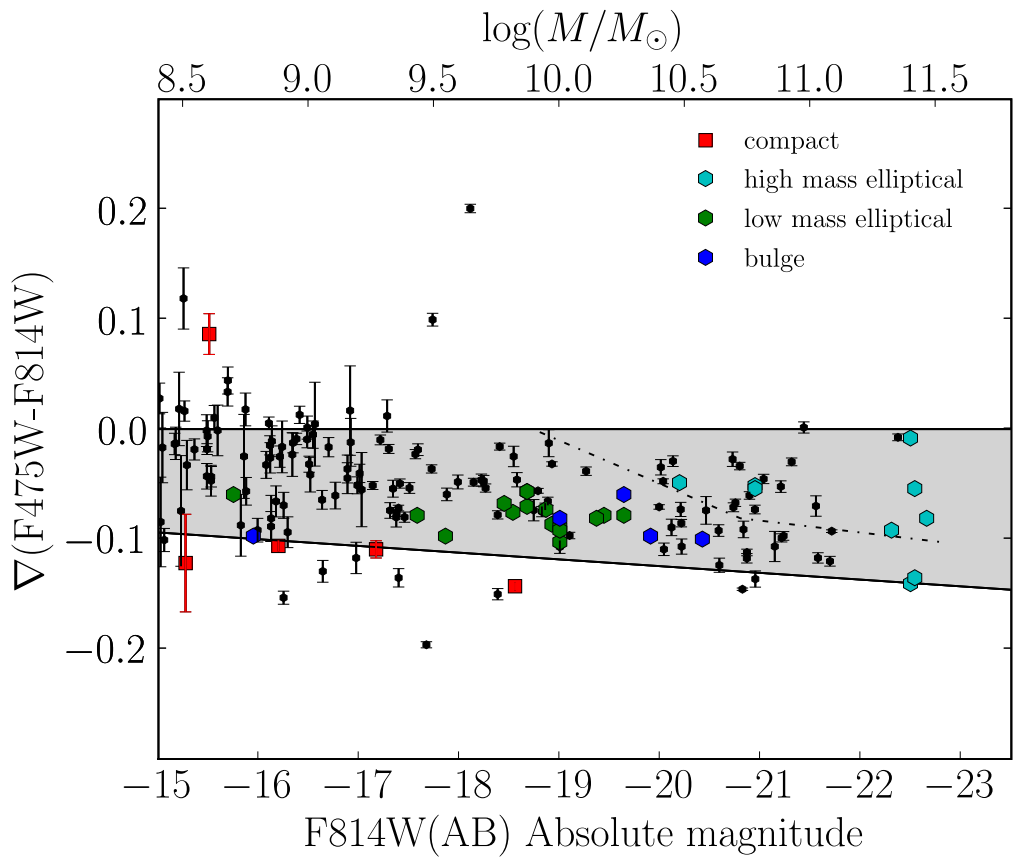}}
\caption{Colour gradients of Coma galaxies (black dots and red squares) as a function of absolute F814W magnitude (and with an approximation of stellar mass) with the models from Pipino et al.(2010) superimposed. Simulated massive ellipticals, low-mass ellipticals and bulges are plotted with cyan, green and blue hexagons, respectively. The grey zone defines metallicity gradients between Pipino et al.'s maximum star formation efficiency-gradients and no gradient.}\label{fig:pipino}
\end{minipage}
\end{figure*}

\subsection{Structural parameters}
In section \ref{chap:struc} we showed that colour gradients also correlated with structural quantities other than magnitude. It is important to know which of these quantities is the primary driver, since it might tell us how the build-up of the galaxy has proceeded. The strongest correlation that we found (even stronger than magnitude) was with S\'ersic index (Spearman rank coefficient 0.6)and as far as we are aware, the strong correlation of colour gradient with S\'ersic index is shown here for the first time. The S\'ersic index is a measure of how centrally concentrated a galaxy is \citep{TruGraCao01}. Although the exact physical processes that determine the S\'ersic index of a galaxy are somewhat clouded, we note that high S\'ersic indices can be produced by violent relaxation \citep[see e.g.][]{van82,HjoMad91}. As colour gradients are to first order a result of dissipational collapse, it is not so clear why they should correlate with a parameter resulting from a non-dissipational process.
Other authors have also looked at correlations with structural parameters. \citet{TorNapCar10} claim to see a dip in colour gradient at S\'ersic index n=2. We do not confirm this with our data (the middle panel of Fig. \ref{fig:grads_struc}), but their data also contain late type galaxies, which may explain this difference.
As a rule of thumb, quantities correlate better with velocity dispersion than with any other structural quantity. All structural quantities shown in this paper are know to correlate with one another at least in some way. It is therefore not so clear if any of these quantities is more fundamental. 
It is however possible that a combination of structural parameters together with the colour gradients form a hyperplane with low scatter, which, when seen in projection against any individual structural parameter, has a consequently larger scatter. To test this, we performed a Principal Component Analysis of three of the structural parameters (magnitude, S\'ersic index $n$, effective radius) and colour gradient. The results are shown in Table \ref{tab:strucpca}. The relative importance of each component is given by the eigenvalue of the component. The most dominant component is obviously the first component. The component divides its energy over the four variables in an almost equal way, indicating that there is no good evidence for a hyperplane, but more for a line. The second most important component shows that colour gradient is inversly related to galaxy size, but this component is not very strong.  
\begin{table}
\begin{tabular}{l|llll}
Eigenvalue & Mag & Eff Radius & S\'ersic Index  & Colour gradient \\
\hline
3.17 & -0.54 & 0.48 & 0.52 & -0.46 \\
0.50 &  0.14 & -0.64 & 0.06 & -0.76 \\
0.25 &  0.23 &  0.54 & -0.67 & -0.46 \\
0.12 & 0.80 & 0.28 & 0.53 & -0.05 \\
\end{tabular}
\caption{Results of a principal components analysis with structural parameters. The first column gives the eigenvalue of each component, the other columns show the relative importance of the observables in this component.}\label{tab:strucpca}
\end{table}
In Figure \ref{fig:grads_struc} we see that especially in the $\nabla$-$\log{r}$ plot, the compact galaxies are outliers. The colours of these sources \citep{PriPhiHux09}, colour gradients and surface brightness at the effective radius (Fig. \ref{fig:grads_struc}) are all similar to those of larger elliptical galaxies, and it appears that only their sizes are different. (The exception here is the smallest compact, which has apparently a positive gradient. However, given the extremely small size of this source, it is quite possible that this is the result of errors in the PSF.) Are the observations of our colour gradients consistent with a tidal threshing scenario? The statistics here are low, but it is tantalizing that the compacts seem to follow a similar $\nabla$-$\log{r}$ relation, which is offset from the 'normal' relation by $\sim 1$ dex in $r$, but we do not see many galaxies in between the two sequences (as a matter of fact, there is an additional galaxy which seems to follow the compacts sequence on the large-end side). Obviously, extrapolating these data back to find the progenitors of these galaxies leads to fairly large and bright galaxies, which would be difficult to strip tidally. It is therefore likely that, if these galaxies were tidally transformed, either additional central star bursts have happened, which made the galaxy metal rich in the centre, or the stripping process managed somehow to enhance an additional gradient. An additional possibility is that we are looking at stripped bulges where still part of the disk is present (note that there is evidence for disc signatures in M32 \citep{Gra02,Set10})  
In Figure \ref{fig:colsersicn} we show how the colour gradient-magnitude relation depends on S\'ersic index, by giving the points a different colour for different S\'ersic indices. This plot gives the impression that in a given magnitude bin, galaxies with higher S\'ersic index have a steeper colour gradient. We warn however that we have not carried out a bulge-disc decomposition for our galaxies and therefore the interpretation of the S\'ersic index may be different for spheroids with disc than for spheroids without disc.  
One of the problems of our sample is that it is a mixed bag of objects. The presence of multiple morphological components in a galaxy impedes a clean interpretation in terms of formation scenarios for elliptical galaxies. As a check, we plot in Fig. \ref{fig:dressler} a set of galaxies for which morphologies have been determined by \citet{Dre80}. Unfortunately, these morphologies are only available for the brightest galaxies. In this figure, we see that in general elliptical galaxies have steeper gradients than S0 galaxies. We can speculate that rotational support has prevented gas from flowing to the centre and therefore thwarted the formation of a strong metallicity gradient. The outlier in this plot is NGC4876, a galaxy with overal quite a flat gradient but a very red centre due to a dust ring. 
\begin{figure}
\begin{minipage}{82mm}
\center
\scalebox{0.41}[0.41]{\includegraphics{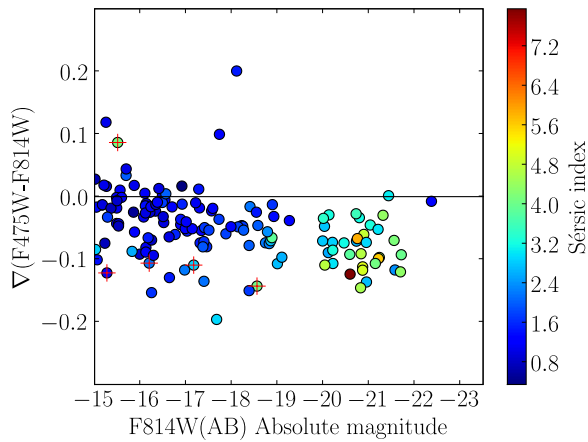}}
\caption{Colour gradients of Coma galaxies of absolute magnitude. The colouring denotes the S\'ersic index of the host galaxy. Interestingly, there is the suggestion that in a given magnitude bin, galaxies with higher S\'ersic index have a steeper colour gradient. Compact galaxies are marked with a red cross.}\label{fig:colsersicn}
\end{minipage}
\end{figure}
\begin{figure}
\begin{minipage}{82mm}
\center
\scalebox{0.41}[0.41]{\includegraphics{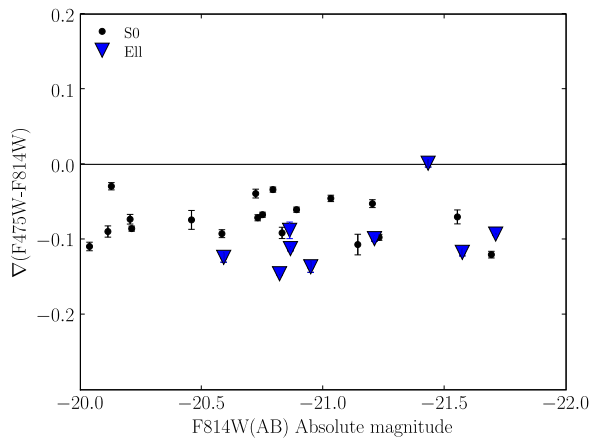}}
\caption{Colour gradients of Coma galaxies for which morphologies are available (from Dressler (1980)), as a function of absolute magnitude. The colouring denotes the S\'ersic index of the host galaxy. Ellipticals have steeper gradients than S0 galaxies. The outlier among the ellipticals is NGC4876.}\label{fig:dressler}
\end{minipage}
\end{figure}

\subsection{The influence of the cluster on gradients}
If the colour gradient of a galaxy is influenced by the star formation efficiency (which in turn is dependent on tidal interactions, pressure in the ICM etc.) we should expect some dependency on clustercentric radius (or at least for the most massive galaxies.) However, since the virial radius of the cluster is $2.9h_{70}^{-1}$ Mpc \citep{okaMam03}, we are only probing the very densest part of the cluster, so a simple explanation may be that the baseline in densities over which we are comparing our gradients is simply not large enough. In addition to this, the fields around NGC 4839 are comparable to the density of a rich group, and it may be the case that the region is sufficiently dense such that star formation has been switched of here just after a first burst. Besides this, we only have a small number of galaxies which we can study. To really investigate the effect of environment, a comparison with galaxies in a much lower density would be the next step. Theoretically one would at least expect that tidal processes which change the orbits of stars to more radial orbits, at least slightly flatten the colour gradient. A comparison with the metallicity gradients from \citet{KoldeRPru09} shows that some of the galaxies in the Fornax cluster (which in terms of density resembles more a rich group than a cluster) have much stronger metallicity gradients than our Coma dwarf galaxies at the same magnitude. This is consistent with the result of \citet{LaBdeCGal05} who found that the steepness of colour gradients is dependent on the richness of the cluster, although the galaxies observed in that paper are much brighter. \refcom{It is possible that there are different formation scenarios for dEs, and that therefore the distribution of colour gradients between our Coma sample and Koleva et al.'s Fornax sample is slightly different.}

\subsection{On the necessity of high resolution data}
Our new result which shows that dwarf galaxies have primarily flat or negative gradients is in contrast with what other authors have found in the past \citep[see e.g.][]{vanBarSki04} and we have shown that this is primarily due to the presence of nuclei and dust discs in the centres of galaxies. High resolution observations have shown that almost all dwarf galaxies have central stellar nuclei, which have different colours than the rest of the galaxy. The fact that \citet{vanBarSki04} detect positive colour gradients in non-nucleated dwarf galaxies has probably more to do with a blue nucleus going unnoticed than that the main component of the galaxy really has a positive colour gradient.
  
It is well known that galaxies can contain more than one component. For elliptical galaxies and dEs, we can often fit a light profile with a single S\'ersic profile, if we exclude the centre, so that we are not plagued by the presence of cores, light excess or nuclear star clusters. However, our data indicates that the colour profile is not always a straight line, but can show breaks. These coincide with the presence of other morphological components, for example, the presence of a nuclear star cluster or a central nuclear disc. Separating the components, and individually studying the photometric properties of them may lead to a better understanding of the formation of these galaxies.   

\section{Summary and Conclusions}
We have derived colour gradients for a sample of confirmed or very likely Coma cluster members. We show that the gradient which one measures is strongly dependent on whether or not one includes the centre of a galaxy. Large ellipticals often have dusty discs in their centre, whereas lower mass elliptical or dwarf galaxies have nuclear star cluster, of which the colour is different from the rest of the galaxy. This shows that for accurate measurement, one really requires high resolution observations.

Our findings indicate that colour gradients in general correlate with galaxy luminosity or mass, in the sense that high-mass galaxies have stronger gradients than dwarfs. However, in general, colour gradients in the main body of the galaxy do not become positive, as has been claimed in the literature.
The colour gradients seem to correlate equally well with other structural parameters, such as effective radius, effective surface brightness and in particular S\'ersic index. Compact galaxies clearly stand out by having steeper gradients than can be expected on the basis of their magnitudes. For a subset of galaxies of which the morphologies are known, we find that S0 galaxies have less steep gradients than elliptical galaxies.
 
We have compared our colour gradients with Mgb gradients and metallicity gradients from the literature, by converting these gradients to colour gradients. The comparison between colour gradients and Mgb gradients is consistent, but the Mgb gradients show generally much larger errorbars. Recently, Spolaor et al. measured an extremely tight relation between metallicity gradient and mass. A comparison of these data with our data shows results consistent with our colour gradients if we include the centre in our fit, but not if we measure the gradient in the outer parts. Since time scales are much longer in the outer parts, and later bursts of star formation primarily influence the centre, the gradient in the outer parts should be a better probe for metallicity gradients from a monolithic collapse scenario. The metallicity gradients from the SAURON data are however fully consistent with our data.

Simulations still have trouble reproducing the observations. Although the work by \citet{PipDErChi10} is able to reproduce the scatter, it apparently misses the trend observed with magnitude. If dwarf galaxies have strong positive age gradients, the results are consistent. 

Given the poor reproduction of gradients by simulations, we can only speculate what our results mean in terms of galaxy formation models. The observed trends with galaxy magnitude imply that somehow the mass of the galaxy, or the potential well are important in shaping the gradient. However, given the scatter in gradients at a given magnitude, this cannot be the only important process. Another clue comes from the S\'ersic index-gradient relation. Galaxies with nearly exponential profiles have flatter gradients than galaxies with higher S\'ersic indices. Higher S\'ersic indices are thought to be the results of processes involving violent relaxation. Since this is a non-dissipational process by itself, the strong gradients may point at a history of moist or wet mergers.   

\section*{Acknowledgments}
Based on observations made with the NASA/ESA {\it Hubble Space Telescope} obtained at the Space Telescope Science Institute, which is operated by the association of Universities for Research in Astronomy, Inc., under NASA contract NAS 5-26555. This research is primarily associated with program GO-10861. We are grateful to the anonymous referee for useful comments. We thank Mina Koleva for providing her Fornax galaxies metallicity data, and Harald Kuntschner for providing his Sauron metallicity gradients. DC and AMK acknowledge support from the Science and Technology Facilities Council, under grant PP/E/001149/1. CH acknowledges financial support from the Estallidos de Formaci\'on Estelar. Fase III, under grant AYA 2007-67965-C03-03/MEC. MB acknowledges financial support from grants AYA2006-12955 and AYA2009-11137 from the Spanish  Ministerio de Ciencia e Innovaci\'on. This research has made use of the NASA/IPAC Extragalactic Database (NED) which is operated by the Jet Propulsion Laboratory, California Institute of Technology, under contract with the National Aeronautics and Space Administration. 

\bibliographystyle{mn2e}
\bibliography{denbrok_gradients}

\begin{thebibliography}{}

\bibitem[\protect\citeauthoryear{{Arimoto} \& {Yoshii}}{{Arimoto} \&
  {Yoshii}}{1987}]{AriYos87}
{Arimoto} N.,  {Yoshii} Y.,  1987, \aap, 173, 23

\bibitem[\protect\citeauthoryear{{Battaglia}, {Tolstoy}, {Helmi}, {Irwin},
  {Letarte}, {Jablonka}, {Hill}, {Venn}, {Shetrone}, {Arimoto}, {Primas},
  {Kaufer}, {Francois}, {Szeifert}, {Abel} \& {Sadakane}}{{Battaglia}
  et~al.}{2006}]{BatTolHel06}
{Battaglia} G.,  {Tolstoy} E.,  {Helmi} A.,  {Irwin} M.~J.,  {Letarte} B.,
  {Jablonka} P.,  {Hill} V.,  {Venn} K.~A.,  {Shetrone} M.~D.,  {Arimoto} N.,
  {Primas} F.,  {Kaufer} A.,  {Francois} P.,  {Szeifert} T.,  {Abel} T.,
  {Sadakane} K.,  2006, \aap, 459, 423

\bibitem[\protect\citeauthoryear{{Bruzual} \& {Charlot}}{{Bruzual} \&
  {Charlot}}{2003}]{BruCha03}
{Bruzual} G.,  {Charlot} S.,  2003, \mnras, 344, 1000

\bibitem[\protect\citeauthoryear{{Carlberg}}{{Carlberg}}{1984}]{Car84}
{Carlberg} R.~G.,  1984, \apj, 286, 403

\bibitem[\protect\citeauthoryear{{Carter}, {Goudfrooij}, {Mobasher} \&
  {Ferguson} H.~C.}{{Carter} et~al.}{2008}]{CarGouMob08}
{Carter} D.,  {Goudfrooij} P.,  {Mobasher} B.,    {Ferguson} H.~C. e.~a.,
  2008, \apjs, 176, 424

\bibitem[\protect\citeauthoryear{{Chiboucas}, {Tully}, {Marzke}, {Trentham},
  {Ferguson}, {Hammer}, {Carter} \& {Khosroshahi}}{{Chiboucas}
  et~al.}{2010}]{ChiTulMar10}
{Chiboucas} K.,  {Tully} R.~B.,  {Marzke} R.~O.,  {Trentham} N.,  {Ferguson}
  H.~C.,  {Hammer} D.,  {Carter} D.,    {Khosroshahi} H.,  2010, \apj, 723, 251

\bibitem[\protect\citeauthoryear{{Chiosi} \& {Carraro}}{{Chiosi} \&
  {Carraro}}{2002}]{ChiCar02}
{Chiosi} C.,  {Carraro} G.,  2002, \mnras, 335, 335

\bibitem[\protect\citeauthoryear{{C{\^o}t{\'e}}, {Piatek}, {Ferrarese},
  {Jord{\'a}n}, {Merritt}, {Peng}, {Ha{\c s}egan}, {Blakeslee}, {Mei}, {West},
  {Milosavljevi{\'c}} \& {Tonry}}{{C{\^o}t{\'e}} et~al.}{2006}]{CotPiaFer06}
{C{\^o}t{\'e}} P.,  {Piatek} S.,  {Ferrarese} L.,  {Jord{\'a}n} A.,  {Merritt}
  D.,  {Peng} E.~W.,  {Ha{\c s}egan} M.,  {Blakeslee} J.~P.,  {Mei} S.,  {West}
  M.~J.,  {Milosavljevi{\'c}} M.,    {Tonry} J.~L.,  2006, \apjs, 165, 57

\bibitem[\protect\citeauthoryear{{de Vaucouleurs}, {de Vaucouleurs}, {Corwin}
  Jr., {Buta}, {Paturel} \& {Fouqu{\'e}}}{{de Vaucouleurs}
  et~al.}{1991}]{deVdeVCor91}
{de Vaucouleurs} G.,  {de Vaucouleurs} A.,  {Corwin} Jr. H.~G.,  {Buta} R.~J.,
  {Paturel} G.,    {Fouqu{\'e}} P.,  1991, {Third Reference Catalogue of Bright
  Galaxies.}

\bibitem[\protect\citeauthoryear{{di Matteo}, {Pipino}, {Lehnert}, {Combes} \&
  {Semelin}}{{di Matteo} et~al.}{2009}]{diMPipLeh09}
{di Matteo} P.,  {Pipino} A.,  {Lehnert} M.~D.,  {Combes} F.,    {Semelin} B.,
  2009, \aap, 499, 427

\bibitem[\protect\citeauthoryear{{Dressler}}{{Dressler}}{1980}]{Dre80}
{Dressler} A.,  1980, \apj, 236, 351

\bibitem[\protect\citeauthoryear{{Falco}, {Kurtz}, {Geller}, {Huchra},
  {Peters}, {Berlind}, {Mink}, {Tokarz} \& {Elwell}}{{Falco}
  et~al.}{1999}]{FalKurGel99}
{Falco} E.~E.,  {Kurtz} M.~J.,  {Geller} M.~J.,  {Huchra} J.~P.,  {Peters} J.,
  {Berlind} P.,  {Mink} D.~J.,  {Tokarz} S.~P.,    {Elwell} B.,  1999, \pasp,
  111, 438

\bibitem[\protect\citeauthoryear{{Ferrarese}, {C{\^o}t{\'e}}, {Jord{\'a}n},
  {Peng}, {Blakeslee}, {Piatek}, {Mei}, {Merritt}, {Milosavljevi{\'c}}, {Tonry}
  \& {West}}{{Ferrarese} et~al.}{2006}]{FerCotJor06}
{Ferrarese} L.,  {C{\^o}t{\'e}} P.,  {Jord{\'a}n} A.,  {Peng} E.~W.,
  {Blakeslee} J.~P.,  {Piatek} S.,  {Mei} S.,  {Merritt} D.,
  {Milosavljevi{\'c}} M.,  {Tonry} J.~L.,    {West} M.~J.,  2006, \apjs, 164,
  334

\bibitem[\protect\citeauthoryear{{Franx}, {Illingworth} \& {Heckman}}{{Franx}
  et~al.}{1989}]{FraIllHec89}
{Franx} M.,  {Illingworth} G.,    {Heckman} T.,  1989, \apj, 344, 613

\bibitem[\protect\citeauthoryear{{Frei} \& {Gunn}}{{Frei} \&
  {Gunn}}{1994}]{FreGun94}
{Frei} Z.,  {Gunn} J.~E.,  1994, \aj, 108, 1476

\bibitem[\protect\citeauthoryear{{Gil de Paz}, {Boissier}, {Madore}, {Seibert}
  \& {et al.}}{{Gil de Paz} et~al.}{2007}]{GilBoiMad07}
{Gil de Paz} A.,  {Boissier} S.,  {Madore} B.~F.,  {Seibert} M.,    {et al.}
  2007, \apjs, 173, 185

\bibitem[\protect\citeauthoryear{{Gorgas}, {Pedraz}, {Guzman}, {Cardiel} \&
  {Gonzalez}}{{Gorgas} et~al.}{1997}]{GorPedGuz97}
{Gorgas} J.,  {Pedraz} S.,  {Guzman} R.,  {Cardiel} N.,    {Gonzalez} J.~J.,
  1997, \apjl, 481, L19+

\bibitem[\protect\citeauthoryear{{Graham}}{{Graham}}{2002}]{Gra02}
{Graham} A.~W.,  2002, \apjl, 568, L13

\bibitem[\protect\citeauthoryear{{Graham} \& {Guzm{\'a}n}}{{Graham} \&
  {Guzm{\'a}n}}{2003}]{GraGuz03}
{Graham} A.~W.,  {Guzm{\'a}n} R.,  2003, \aj, 125, 2936

\bibitem[\protect\citeauthoryear{{Gunn} \& {Gott} III}{{Gunn} \&
  {Gott}}{1972}]{GunGot72}
{Gunn} J.~E.,  {Gott} III J.~R.,  1972, \apj, 176, 1

\bibitem[\protect\citeauthoryear{{Hammer}, {Verdoes Kleijn}, {Hoyos}, {den
  Brok}, {Balcells}, {Ferguson}, {Goudfrooij}, {Carter}, {Guzman}, {Peletier},
  {Smith}, {Graham}, {Trentham}, {Peng}, {Puzia}, {Lucey}, {Jogee} \& {et
  al.}}{{Hammer} et~al.}{2010}]{HamVerHoy10}
{Hammer} D.,  {Verdoes Kleijn} G.,  {Hoyos} C.,  {den Brok} M.,  {Balcells} M.,
   {Ferguson} H.~C.,  {Goudfrooij} P.,  {Carter} D.,  {Guzman} R.,  {Peletier}
  R.~F.,  {Smith} R.~J.,  {Graham} A.~W.,  {Trentham} N.,  {Peng} E.,  {Puzia}
  T.~H.,  {Lucey} J.~R.,  {Jogee} S.,    {et al.} 2010, ArXiv e-prints

\bibitem[\protect\citeauthoryear{{Hjorth} \& {Madsen}}{{Hjorth} \&
  {Madsen}}{1991}]{HjoMad91}
{Hjorth} J.,  {Madsen} J.,  1991, \mnras, 253, 703

\bibitem[\protect\citeauthoryear{{Hopkins}, {Cox}, {Dutta}, {Hernquist},
  {Kormendy} \& {Lauer}}{{Hopkins} et~al.}{2009}]{HopCoxDut09}
{Hopkins} P.~F.,  {Cox} T.~J.,  {Dutta} S.~N.,  {Hernquist} L.,  {Kormendy} J.,
     {Lauer} T.~R.,  2009, \apjs, 181, 135

\bibitem[\protect\citeauthoryear{{Hoyos}, {den Brok}, {Verdoes Kleijn},
  {Carter}, {Balcells}, {Guzman}, {Peletier}, {Ferguson}, {Goudfrooij},
  {Graham}, {Hammer}, {Karick}, {Lucey}, {Matkovic}, {Merritt}, {Mouhcine} \&
  {Valentijn}}{{Hoyos} et~al.}{2010}]{HoydenVer10}
{Hoyos} C.,  {den Brok} M.,  {Verdoes Kleijn} G.,  {Carter} D.,  {Balcells} M.,
   {Guzman} R.,  {Peletier} R.,  {Ferguson} H.~C.,  {Goudfrooij} P.,  {Graham}
  A.~W.,  {Hammer} D.,  {Karick} A.~M.,  {Lucey} J.~R.,  {Matkovic} A.,
  {Merritt} D.,  {Mouhcine} M.,    {Valentijn} E.,  2010, ArXiv e-prints

\bibitem[\protect\citeauthoryear{{Jord{\'a}n}, {Blakeslee}, {C{\^o}t{\'e}},
  {Ferrarese}, {Infante}, {Mei}, {Merritt}, {Peng}, {Tonry} \&
  {West}}{{Jord{\'a}n} et~al.}{2007}]{JorBlaCot07}
{Jord{\'a}n} A.,  {Blakeslee} J.~P.,  {C{\^o}t{\'e}} P.,  {Ferrarese} L.,
  {Infante} L.,  {Mei} S.,  {Merritt} D.,  {Peng} E.~W.,  {Tonry} J.~L.,
  {West} M.~J.,  2007, \apjs, 169, 213

\bibitem[\protect\citeauthoryear{{J{\o}rgensen}, {Franx} \&
  {Kj{\ae}rgaard}}{{J{\o}rgensen} et~al.}{1992}]{JrgFraKjr92}
{J{\o}rgensen} I.,  {Franx} M.,    {Kj{\ae}rgaard} P.,  1992, \aaps, 95, 489

\bibitem[\protect\citeauthoryear{{Kawata} \& {Gibson}}{{Kawata} \&
  {Gibson}}{2003}]{KawGib03}
{Kawata} D.,  {Gibson} B.~K.,  2003, \mnras, 340, 908

\bibitem[\protect\citeauthoryear{{Kobayashi}}{{Kobayashi}}{2004}]{Kob04}
{Kobayashi} C.,  2004, \mnras, 347, 740

\bibitem[\protect\citeauthoryear{{Koleva}, {de Rijcke}, {Prugniel}, {Zeilinger}
  \& {Michielsen}}{{Koleva} et~al.}{2009}]{KoldeRPru09}
{Koleva} M.,  {de Rijcke} S.,  {Prugniel} P.,  {Zeilinger} W.~W.,
  {Michielsen} D.,  2009, \mnras, 396, 2133

\bibitem[\protect\citeauthoryear{{Kormendy}}{{Kormendy}}{1985}]{Kor85}
{Kormendy} J.,  1985, \apj, 295, 73

\bibitem[\protect\citeauthoryear{{Kuntschner}, {Emsellem}, {Bacon},
  {Cappellari}, {Davies}, {de Zeeuw}, {Falc{\'o}n-Barroso}, {Krajnovi{\'c}},
  {McDermid}, {Peletier}, {Sarzi}, {Shapiro}, {van den Bosch} \& {van de
  Ven}}{{Kuntschner} et~al.}{2010}]{KunEmsBac10}
{Kuntschner} H.,  {Emsellem} E.,  {Bacon} R.,  {Cappellari} M.,  {Davies}
  R.~L.,  {de Zeeuw} P.~T.,  {Falc{\'o}n-Barroso} J.,  {Krajnovi{\'c}} D.,
  {McDermid} R.~M.,  {Peletier} R.~F.,  {Sarzi} M.,  {Shapiro} K.~L.,  {van den
  Bosch} R.~C.~E.,    {van de Ven} G.,  2010, ArXiv e-prints

\bibitem[\protect\citeauthoryear{{La Barbera} \& {de Carvalho}}{{La Barbera} \&
  {de Carvalho}}{2009}]{LaBdeC09}
{La Barbera} F.,  {de Carvalho} R.~R.,  2009, \apjl, 699, L76

\bibitem[\protect\citeauthoryear{{La Barbera}, {de Carvalho}, {Gal},
  {Busarello}, {Merluzzi}, {Capaccioli} \& {Djorgovski}}{{La Barbera}
  et~al.}{2005}]{LaBdeCGal05}
{La Barbera} F.,  {de Carvalho} R.~R.,  {Gal} R.~R.,  {Busarello} G.,
  {Merluzzi} P.,  {Capaccioli} M.,    {Djorgovski} S.~G.,  2005, \apjl, 626,
  L19

\bibitem[\protect\citeauthoryear{{La Barbera}, {Merluzzi}, {Busarello},
  {Massarotti} \& {Mercurio}}{{La Barbera} et~al.}{2004}]{LaBMerBus04}
{La Barbera} F.,  {Merluzzi} P.,  {Busarello} G.,  {Massarotti} M.,
  {Mercurio} A.,  2004, \aap, 425, 797

\bibitem[\protect\citeauthoryear{{Larson}}{{Larson}}{1974}]{Lar74}
{Larson} R.~B.,  1974, \mnras, 166, 585

\bibitem[\protect\citeauthoryear{{Lauer}, {Faber}, {Ajhar}, {Grillmair} \&
  {Scowen}}{{Lauer} et~al.}{1998}]{LauFabAjh98}
{Lauer} T.~R.,  {Faber} S.~M.,  {Ajhar} E.~A.,  {Grillmair} C.~J.,    {Scowen}
  P.~A.,  1998, \aj, 116, 2263

\bibitem[\protect\citeauthoryear{{Lianou}, {Grebel} \& {Koch}}{{Lianou}
  et~al.}{2010}]{LiaGreKoc10}
{Lianou} S.,  {Grebel} E.~K.,    {Koch} A.,  2010, ArXiv e-prints

\bibitem[\protect\citeauthoryear{{{\L}okas} \& {Mamon}}{{{\L}okas} \&
  {Mamon}}{2003}]{okaMam03}
{{\L}okas} E.~L.,  {Mamon} G.~A.,  2003, \mnras, 343, 401

\bibitem[\protect\citeauthoryear{{Mehlert}, {Thomas}, {Saglia}, {Bender} \&
  {Wegner}}{{Mehlert} et~al.}{2003}]{MehThoSag03}
{Mehlert} D.,  {Thomas} D.,  {Saglia} R.~P.,  {Bender} R.,    {Wegner} G.,
  2003, \aap, 407, 423

\bibitem[\protect\citeauthoryear{{Moore}, {Katz}, {Lake}, {Dressler} \&
  {Oemler}}{{Moore} et~al.}{1996}]{MooKatLak96}
{Moore} B.,  {Katz} N.,  {Lake} G.,  {Dressler} A.,    {Oemler} A.,  1996,
  \nat, 379, 613

\bibitem[\protect\citeauthoryear{{Ogando}, {Maia}, {Pellegrini} \& {da
  Costa}}{{Ogando} et~al.}{2008}]{OgaMaiPel08}
{Ogando} R.~L.~C.,  {Maia} M.~A.~G.,  {Pellegrini} P.~S.,    {da Costa} L.~N.,
  2008, \aj, 135, 2424

\bibitem[\protect\citeauthoryear{{Peletier}}{{Peletier}}{1993}]{Pel93}
{Peletier} R.~F.,  1993, \aap, 271, 51

\bibitem[\protect\citeauthoryear{{Peletier}, {Davies}, {Illingworth}, {Davis}
  \& {Cawson}}{{Peletier} et~al.}{1990}]{PelDavIll90}
{Peletier} R.~F.,  {Davies} R.~L.,  {Illingworth} G.~D.,  {Davis} L.~E.,
  {Cawson} M.,  1990, \aj, 100, 1091

\bibitem[\protect\citeauthoryear{{Peletier}, {Valentijn} \&
  {Jameson}}{{Peletier} et~al.}{1990}]{PelValJam90}
{Peletier} R.~F.,  {Valentijn} E.~A.,    {Jameson} R.~F.,  1990, \aap, 233, 62

\bibitem[\protect\citeauthoryear{{Pipino}, {D'Ercole}, {Chiappini} \&
  {Matteucci}}{{Pipino} et~al.}{2010}]{PipDErChi10}
{Pipino} A.,  {D'Ercole} A.,  {Chiappini} C.,    {Matteucci} F.,  2010, ArXiv
  e-prints

\bibitem[\protect\citeauthoryear{{Pipino}, {D'Ercole} \& {Matteucci}}{{Pipino}
  et~al.}{2008}]{PipDErMat08}
{Pipino} A.,  {D'Ercole} A.,    {Matteucci} F.,  2008, \aap, 484, 679

\bibitem[\protect\citeauthoryear{{Price}, {Phillipps}, {Huxor}, {Trentham},
  {Ferguson}, {Marzke}, {Hornschemeier}, {Goudfrooij}, {Hammer}, {Tully},
  {Chiboucas}, {Smith}, {Carter}, {Merritt}, {Balcells}, {Erwin} \&
  {Puzia}}{{Price} et~al.}{2009}]{PriPhiHux09}
{Price} J.,  {Phillipps} S.,  {Huxor} A.,  {Trentham} N.,  {Ferguson} H.~C.,
  {Marzke} R.~O.,  {Hornschemeier} A.,  {Goudfrooij} P.,  {Hammer} D.,  {Tully}
  R.~B.,  {Chiboucas} K.,  {Smith} R.~J.,  {Carter} D.,  {Merritt} D.,
  {Balcells} M.,  {Erwin} P.,    {Puzia} T.~H.,  2009, \mnras, 397, 1816

\bibitem[\protect\citeauthoryear{{Rawle}, {Smith} \& {Lucey}}{{Rawle}
  et~al.}{2010}]{RawSmiLuc10}
{Rawle} T.~D.,  {Smith} R.~J.,    {Lucey} J.~R.,  2010, \mnras, 401, 852

\bibitem[\protect\citeauthoryear{{Rose}, {Arimoto}, {Caldwell}, {Schiavon},
  {Vazdekis} \& {Yamada}}{{Rose} et~al.}{2005}]{RosAriCal05}
{Rose} J.~A.,  {Arimoto} N.,  {Caldwell} N.,  {Schiavon} R.~P.,  {Vazdekis} A.,
     {Yamada} Y.,  2005, \aj, 129, 712

\bibitem[\protect\citeauthoryear{{Saglia}, {Maraston}, {Greggio}, {Bender} \&
  {Ziegler}}{{Saglia} et~al.}{2000}]{SagMarGre00}
{Saglia} R.~P.,  {Maraston} C.,  {Greggio} L.,  {Bender} R.,    {Ziegler} B.,
  2000, \aap, 360, 911

\bibitem[\protect\citeauthoryear{{S{\'a}nchez-Bl{\'a}zquez}, {Gorgas} \&
  {Cardiel}}{{S{\'a}nchez-Bl{\'a}zquez} et~al.}{2006}]{SanGorCar06}
{S{\'a}nchez-Bl{\'a}zquez} P.,  {Gorgas} J.,    {Cardiel} N.,  2006, \aap, 457,
  823

\bibitem[\protect\citeauthoryear{{Sandage}}{{Sandage}}{1972}]{San72}
{Sandage} A.,  1972, \apj, 176, 21

\bibitem[\protect\citeauthoryear{{Schlegel}, {Finkbeiner} \&
  {Davis}}{{Schlegel} et~al.}{1998}]{SchFinDav98}
{Schlegel} D.~J.,  {Finkbeiner} D.~P.,    {Davis} M.,  1998, \apj, 500, 525

\bibitem[\protect\citeauthoryear{{Seth}}{{Seth}}{2010}]{Set10}
{Seth} A.~C.,  2010, ArXiv e-prints

\bibitem[\protect\citeauthoryear{{Sirianni}, {Jee}, {Ben{\'{\i}}tez},
  {Blakeslee}, {Martel}, {Meurer}, {Clampin}, {De Marchi}, {Ford}, {Gilliland},
  {Hartig}, {Illingworth}, {Mack} \& {McCann}}{{Sirianni}
  et~al.}{2005}]{SirJeeBen05}
{Sirianni} M.,  {Jee} M.~J.,  {Ben{\'{\i}}tez} N.,  {Blakeslee} J.~P.,
  {Martel} A.~R.,  {Meurer} G.,  {Clampin} M.,  {De Marchi} G.,  {Ford} H.~C.,
  {Gilliland} R.,  {Hartig} G.~F.,  {Illingworth} G.~D.,  {Mack} J.,
  {McCann} W.~J.,  2005, \pasp, 117, 1049

\bibitem[\protect\citeauthoryear{{Smith}, {Lucey}, {Hudson}, {Allanson},
  {Bridges}, {Hornschemeier}, {Marzke} \& {Miller}}{{Smith}
  et~al.}{2009}]{SmiLucHud09}
{Smith} R.~J.,  {Lucey} J.~R.,  {Hudson} M.~J.,  {Allanson} S.~P.,  {Bridges}
  T.~J.,  {Hornschemeier} A.~E.,  {Marzke} R.~O.,    {Miller} N.~A.,  2009,
  \mnras, 392, 1265

\bibitem[\protect\citeauthoryear{{Smith}, {Marzke}, {Hornschemeier}, {Bridges},
  {Hudson}, {Miller}, {Lucey}, {V{\'a}zquez} \& {Carter}}{{Smith}
  et~al.}{2008}]{SmiMarHor08}
{Smith} R.~J.,  {Marzke} R.~O.,  {Hornschemeier} A.~E.,  {Bridges} T.~J.,
  {Hudson} M.~J.,  {Miller} N.~A.,  {Lucey} J.~R.,  {V{\'a}zquez} G.~A.,
  {Carter} D.,  2008, \mnras, 386, L96

\bibitem[\protect\citeauthoryear{{Spolaor}, {Kobayashi}, {Forbes}, {Couch} \&
  {Hau}}{{Spolaor} et~al.}{2010}]{SpoKobFor10}
{Spolaor} M.,  {Kobayashi} C.,  {Forbes} D.~A.,  {Couch} W.~J.,    {Hau}
  G.~K.~T.,  2010, ArXiv e-prints

\bibitem[\protect\citeauthoryear{{Spolaor}, {Proctor}, {Forbes} \&
  {Couch}}{{Spolaor} et~al.}{2009}]{SpoProFor09}
{Spolaor} M.,  {Proctor} R.~N.,  {Forbes} D.~A.,    {Couch} W.~J.,  2009,
  \apjl, 691, L138

\bibitem[\protect\citeauthoryear{{Stinson}, {Dalcanton}, {Quinn}, {Gogarten},
  {Kaufmann} \& {Wadsley}}{{Stinson} et~al.}{2009}]{StiDalQui09}
{Stinson} G.~S.,  {Dalcanton} J.~J.,  {Quinn} T.,  {Gogarten} S.~M.,
  {Kaufmann} T.,    {Wadsley} J.,  2009, \mnras, 395, 1455

\bibitem[\protect\citeauthoryear{{Thomas}, {Maraston} \& {Bender}}{{Thomas}
  et~al.}{2003}]{ThoMarBen03}
{Thomas} D.,  {Maraston} C.,    {Bender} R.,  2003, \mnras, 339, 897

\bibitem[\protect\citeauthoryear{{Thomas}, {Maraston} \& {Korn}}{{Thomas}
  et~al.}{2004}]{ThoMarKor04}
{Thomas} D.,  {Maraston} C.,    {Korn} A.,  2004, \mnras, 351, L19

\bibitem[\protect\citeauthoryear{{Tortora}, {Napolitano}, {Cardone},
  {Capaccioli}, {Jetzer} \& {Molinaro}}{{Tortora} et~al.}{2010}]{TorNapCar10}
{Tortora} C.,  {Napolitano} N.~R.,  {Cardone} V.~F.,  {Capaccioli} M.,
  {Jetzer} P.,    {Molinaro} R.,  2010, ArXiv e-prints

\bibitem[\protect\citeauthoryear{{Trujillo}, {Graham} \& {Caon}}{{Trujillo}
  et~al.}{2001}]{TruGraCao01}
{Trujillo} I.,  {Graham} A.~W.,    {Caon} N.,  2001, \mnras, 326, 869

\bibitem[\protect\citeauthoryear{{Tully} \& {Trentham}}{{Tully} \&
  {Trentham}}{2008}]{TulTre08}
{Tully} R.~B.,  {Trentham} N.,  2008, \aj, 135, 1488

\bibitem[\protect\citeauthoryear{{Vader}, {Vigroux}, {Lachieze-Rey} \&
  {Souviron}}{{Vader} et~al.}{1988}]{VadVigLac88}
{Vader} J.~P.,  {Vigroux} L.,  {Lachieze-Rey} M.,    {Souviron} J.,  1988,
  \aap, 203, 217

\bibitem[\protect\citeauthoryear{{Valcke}, {de Rijcke} \& {Dejonghe}}{{Valcke}
  et~al.}{2008}]{ValdeRDej08}
{Valcke} S.,  {de Rijcke} S.,    {Dejonghe} H.,  2008, \mnras, 389, 1111

\bibitem[\protect\citeauthoryear{{van Albada}}{{van Albada}}{1982}]{van82}
{van Albada} T.~S.,  1982, \mnras, 201, 939

\bibitem[\protect\citeauthoryear{{van Zee}, {Barton} \& {Skillman}}{{van Zee}
  et~al.}{2004}]{vanBarSki04}
{van Zee} L.,  {Barton} E.~J.,    {Skillman} E.~D.,  2004, \aj, 128, 2797

\bibitem[\protect\citeauthoryear{{Vazdekis}, {S{\'a}nchez-Bl{\'a}zquez},
  {Falc{\'o}n-Barroso}, {Cenarro}, {Beasley}, {Cardiel}, {Gorgas} \&
  {Peletier}}{{Vazdekis} et~al.}{2010}]{VazSanFal10}
{Vazdekis} A.,  {S{\'a}nchez-Bl{\'a}zquez} P.,  {Falc{\'o}n-Barroso} J.,
  {Cenarro} A.~J.,  {Beasley} M.~A.,  {Cardiel} N.,  {Gorgas} J.,    {Peletier}
  R.~F.,  2010, \mnras, 404, 1639

\bibitem[\protect\citeauthoryear{{White}}{{White}}{1980}]{Whi80}
{White} S.~D.~M.,  1980, \mnras, 191, 1P

\end{thebibliography}
\appendix
\onecolumn
\section{}
\begin{longtable}{lllcccccc}
\hline
COMA ID & $\alpha$(J2000) & $\delta$(J2000) & F814W & rating &redshift & gradient & $\epsilon$ & $\chi_{\nu}$ \\
\hline
\hline
COMAi125635.495p271430.37 &  194.14789 & 27.24176 & 17.79 & 0 & 0.0187 & -0.010$\pm$0.006 & 0.33 & 0.91 \\
COMAi125636.641p271504.06 &  194.15266 & 27.25111 & 19.51 & 2 &  ...  & -0.007$\pm$0.009 & 0.32 & 2.61 \\
COMAi125636.787p271247.99 &  194.15329 & 27.21331 & 17.28 & 2 &  ...  & -0.036$\pm$0.007 & 0.02 & 5.54 \\
COMAi125704.336p273133.26 &  194.26808 & 27.52592 & 15.74 & 0 & 0.0277 & -0.038$\pm$0.000 & 0.36 & 0.89 \\
COMAi125708.355p272924.03 &  194.28477 & 27.49009 & 19.72 & 0 & 0.0252 & -0.033$\pm$0.049 & 0.03 & 0.44 \\
COMAi125710.767p272417.44 &  194.29485 & 27.40484 & 14.14 & 0 & 0.0206 & -0.117$\pm$0.002 & 0.03 & 5.23 \\
COMAi125711.016p273142.35 &  194.29593 & 27.52845 & 18.62 & 1 &  ...  & -0.009$\pm$0.008 & 0.33 & 1.43 \\
COMAi125712.264p272313.35 &  194.30106 & 27.38704 & 18.66 & 2 &  ...  & -0.013$\pm$0.010 & 0.32 & 1.77 \\
COMAi125713.240p272437.24 &  194.30515 & 27.41036 & 18.37 & 0 & 0.0245 & -0.064$\pm$0.015 & 0.18 & 0.74 \\
COMAi125815.275p272752.96 &  194.56358 & 27.46484 & 17.63 & 0 & 0.0254 & -0.080$\pm$0.005 & 0.53 & 0.70 \\
COMAi125820.533p272546.03 &  194.58560 & 27.42943 & 16.46 & 0 & 0.0251 & -0.025$\pm$0.004 & 0.42 & 0.25 \\
COMAi125828.358p271315.01 &  194.61818 & 27.22083 & 18.59 & 0 & 0.0247 & 0.013$\pm$0.007 & 0.22 & 0.82 \\
COMAi125832.060p272722.85 &  194.63365 & 27.45627 & 14.42 & 0 & 0.0234 & -0.092$\pm$0.004 & 0.34 & 0.49 \\
COMAi125845.544p274513.68 &  194.68977 & 27.75379 & 16.62 & 0 & 0.0222 & -0.078$\pm$0.008 & 0.08 & 1.06 \\
COMAi125852.097p274706.14 &  194.71709 & 27.78505 & 14.06 & 0 & 0.0189 & -0.073$\pm$0.004 & 0.25 & 0.33 \\
COMAi125853.079p274741.66 &  194.72118 & 27.79493 & 19.31 & 2 &  ...  & 0.044$\pm$0.019 & 0.05 & 1.11 \\
COMAi125856.954p274719.83 &  194.73732 & 27.78885 & 19.99 & 2 & 0.0229 & 0.028$\pm$0.015 & 0.30 & 0.70 \\
COMAi125902.433p28021.36 &  194.76012 & 28.00591 & 18.93 & 0 & 0.0266 & -0.033$\pm$0.013 & 0.12 & 1.01 \\
COMAi125904.792p28301.21 &  194.76994 & 28.05034 & 16.18 & 0 & 0.0267 & -0.075$\pm$0.002 & 0.35 & 0.80 \\
COMAi125905.938p28228.86 &  194.77472 & 28.04135 & 18.03 & 2 &  ...  & -0.117$\pm$0.020 & 0.03 & 1.44 \\
COMAi125909.465p28227.38 &  194.78941 & 28.04094 & 16.00 & 0 & 0.0240 & -0.107$\pm$0.003 & 0.51 & 0.23 \\
COMAi125911.545p28033.38 &  194.79808 & 28.00926 & 16.62 & 0 & 0.0230 & -0.150$\pm$0.003 & 0.32 & 1.84 \\
COMAi125920.904p28057.49 &  194.83708 & 28.01591 & 19.94 & 0 & 0.0227 & -0.101$\pm$0.018 & 0.13 & 2.60 \\
COMAi125921.615p28102.21 &  194.84004 & 28.01723 & 18.75 & 0 & 0.0222 & -0.069$\pm$0.012 & 0.39 & 0.49 \\
COMAi125926.459p275124.76 &  194.86024 & 27.85684 & 17.50 & 0 & 0.0166 & -0.054$\pm$0.012 & 0.13 & 1.01 \\
COMAi125927.221p275257.00 &  194.86340 & 27.88250 & 19.41 & 0 & 0.0225 & -0.001$\pm$0.045 & 0.06 & 0.69 \\
COMAi125927.694p28145.91 &  194.86540 & 28.02939 & 19.52 & 0 & 0.0260 & -0.043$\pm$0.025 & 0.07 & 1.22 \\
COMAi125928.503p28109.38 &  194.86876 & 28.01924 & 17.86 & 0 & 0.0200 & -0.051$\pm$0.007 & 0.02 & 2.29 \\
COMAi125928.728p28225.90 &  194.86972 & 28.04050 & 14.87 & 0 & 0.0188 & -0.029$\pm$0.002 & 0.38 & 0.73 \\
COMAi125929.404p275100.51 &  194.87253 & 27.85010 & 13.86 & 0 & 0.0227 & -0.107$\pm$0.010 & 0.32 & 0.10 \\
COMAi125929.956p275723.26 &  194.87485 & 27.95648 & 13.45 & 0 & 0.0226 & -0.070$\pm$0.015 & 0.14 & 0.30 \\
COMAi125930.268p28115.17 &  194.87611 & 28.02085 & 17.55 & 0 & 0.0240 & -0.080$\pm$0.005 & 0.31 & 1.97 \\
COMAi125930.825p275303.42 &  194.87840 & 27.88429 & 13.80 & 0 & 0.0157 & -0.052$\pm$0.004 & 0.18 & 0.79 \\
COMAi125930.828p28230.68 &  194.87848 & 28.04184 & 18.12 & 0 & 0.0218 & -0.036$\pm$0.033 & 0.07 & 1.20 \\
COMAi125931.115p275717.73 &  194.87968 & 27.95494 & 16.11 & 0 & 0.0236 & -0.013$\pm$0.027 & 0.34 & 0.16 \\
COMAi125931.455p28247.62 &  194.88109 & 28.04654 & 14.27 & 0 & 0.0230 & -0.071$\pm$0.004 & 0.27 & 0.38 \\
COMAi125932.789p275900.95 &  194.88663 & 27.98362 & 13.31 & 0 & 0.0194 & -0.120$\pm$0.005 & 0.13 & 2.50 \\
COMAi125933.235p28152.56 &  194.88849 & 28.03125 & 18.50 & 0 & 0.0237 & -0.032$\pm$0.015 & 0.02 & 1.08 \\
COMAi125933.270p28152.42 &  194.88852 & 28.03129 & 18.67 & 0 & 0.0237 & -0.023$\pm$0.019 & 0.11 & 0.24 \\
COMAi125934.360p275943.32 &  194.89316 & 27.99539 & 19.83 & 0 & 0.0176 & -0.013$\pm$0.021 & 0.13 & 2.75 \\
COMAi125935.286p275149.16 &  194.89700 & 27.86365 & 16.08 & 0 & 0.0209 & -0.032$\pm$0.005 & 0.13 & 1.09 \\
COMAi125935.698p275733.36 &  194.89876 & 27.95926 & 12.63 & 0 & 0.0239 & -0.007$\pm$0.009 & 0.03 & 0.24 \\
COMAi125936.461p275107.46 &  194.90190 & 27.85205 & 19.75 & 2 &  ...  & 0.119$\pm$0.027 & 0.48 & 0.89 \\
COMAi125937.010p28106.95 &  194.90422 & 28.01862 & 17.59 & 0 & 0.0240 & -0.049$\pm$0.009 & 0.15 & 1.51 \\
COMAi125937.200p275819.97 &  194.90500 & 27.97220 & 17.83 & 0 & 0.0254 & -0.109$\pm$0.024 & 0.11 & 0.27 \\
COMAi125937.208p275213.73 &  194.90500 & 27.87049 & 17.61 & 0 & 0.0211 & -0.072$\pm$0.003 & 0.04 & 0.74 \\
COMAi125937.988p28003.56 &  194.90825 & 28.00100 & 15.90 & 0 & 0.0165 & -0.097$\pm$0.003 & 0.07 & 1.21 \\
COMAi125938.323p275913.84 &  194.90965 & 27.98718 & 14.79 & 0 & 0.0227 & -0.086$\pm$0.008 & 0.27 & 1.96 \\
COMAi125939.218p275954.67 &  194.91338 & 27.99852 & 19.49 & 0 & 0.0236 & 0.086$\pm$0.028 & 0.10 & 0.45 \\
COMAi125939.657p275713.86 &  194.91524 & 27.95381 & 14.99 & 0 & 0.0268 & -0.035$\pm$0.003 & 0.32 & 0.25 \\
COMAi125940.278p275805.73 &  194.91782 & 27.96823 & 14.96 & 0 & 0.0253 & -0.109$\pm$0.012 & 0.30 & 0.85 \\
COMAi125942.306p275529.11 &  194.92625 & 27.92475 & 14.41 & 0 & 0.0231 & -0.124$\pm$0.009 & 0.17 & 0.32 \\
COMAi125942.373p28158.52 &  194.92654 & 28.03292 & 18.30 & 0 & 0.0263 & -0.016$\pm$0.022 & 0.07 & 0.62 \\
COMAi125942.888p28202.24 &  194.92870 & 28.03395 & 18.79 & 2 &  ...  & -0.025$\pm$0.018 & 0.09 & 0.82 \\
COMAi125943.540p275620.63 &  194.93140 & 27.93906 & 18.77 & 0 & 0.0236 & -0.016$\pm$0.044 & 0.04 & 0.35 \\
COMAi125943.724p275940.89 &  194.93214 & 27.99467 & 14.18 & 0 & 0.0223 & -0.146$\pm$0.002 & 0.01 & 11.49 \\
COMAi125944.182p275730.39 &  194.93407 & 27.95839 & 14.79 & 0 & 0.0230 & -0.107$\pm$0.007 & 0.45 & 0.14 \\
COMAi125944.217p275730.29 &  194.93425 & 27.95851 & 14.80 & 0 & 0.0230 & -0.073$\pm$0.009 & 0.44 & 0.62 \\
COMAi125944.404p275444.93 &  194.93500 & 27.91248 & 13.57 & 0 & 0.0224 & 0.002$\pm$0.004 & 0.22 & 1.19 \\
COMAi125946.701p28000.43 &  194.94463 & 28.00012 & 17.70 & 2 &  ...  & -0.018$\pm$0.010 & 0.04 & 1.11 \\
COMAi125946.775p275825.88 &  194.94493 & 27.97391 & 13.69 & 0 & 0.0313 & -0.030$\pm$0.002 & 0.20 & 0.44 \\
COMAi125946.941p275930.84 &  194.94565 & 27.99193 & 16.86 & 0 & 0.0279 & -0.048$\pm$0.007 & 0.17 & 1.26 \\
COMAi125947.675p275425.89 &  194.94864 & 27.90720 & 19.77 & 0 & 0.0284 & -0.074$\pm$0.086 & 0.24 & 0.14 \\
COMAi125948.590p275858.01 &  194.95251 & 27.98281 & 17.44 & 0 & 0.0180 & -0.023$\pm$0.006 & 0.25 & 1.22 \\
COMAi125949.065p275834.01 &  194.95448 & 27.97616 & 19.52 & 0 & 0.0166 & -0.001$\pm$0.029 & 0.04 & 1.79 \\
COMAi125950.181p275445.54 &  194.95909 & 27.91266 & 17.02 & 0 & 0.0243 & -0.048$\pm$0.007 & 0.18 & 0.43 \\
COMAi125950.410p275518.20 &  194.96005 & 27.92174 & 19.15 & 2 &  ...  & -0.025$\pm$0.082 & 0.13 & 1.72 \\
COMAi125951.477p275935.40 &  194.96457 & 27.99317 & 19.47 & 0 & 0.0240 & -0.043$\pm$0.017 & 0.12 & 1.61 \\
COMAi125951.837p275726.21 &  194.96602 & 27.95734 & 19.98 & 0 & 0.0201 & -0.084$\pm$0.056 & 0.12 & 1.78 \\
COMAi125953.930p275813.76 &  194.97475 & 27.97051 & 17.41 & 0 & 0.0225 & -0.019$\pm$0.003 & 0.05 & 1.05 \\
COMAi125955.702p275503.79 &  194.98212 & 27.91774 & 17.69 & 0 & 0.0221 & -0.074$\pm$0.017 & 0.02 & 0.69 \\
COMAi125955.939p275748.80 &  194.98312 & 27.96358 & 19.13 & 0 & 0.0232 & 0.018$\pm$0.031 & 0.18 & 0.56 \\
COMAi125958.220p275410.82 &  194.99251 & 27.90301 & 18.89 & 0 & 0.0244 & -0.026$\pm$0.017 & 0.25 & 0.97 \\
COMAi125959.476p275626.04 &  194.99783 & 27.94059 & 16.22 & 0 & 0.0205 & -0.056$\pm$0.005 & 0.05 & 2.08 \\
COMAi125959.901p275921.80 &  194.99966 & 27.98935 & 19.73 & 0 & 0.0218 & -0.122$\pm$0.092 & 0.02 & 0.05 \\
COMAi13000.946p275643.80 &  195.00397 & 27.94552 & 16.12 & 0 & 0.0228 & -0.066$\pm$0.012 & 0.11 & 0.72 \\
COMAi13000.986p275929.72 &  195.00420 & 27.99154 & 17.72 & 0 & 0.0251 & 0.012$\pm$0.037 & 0.03 & 0.16 \\
COMAi13004.048p275342.93 &  195.01677 & 27.89528 & 19.18 & 2 &  ...  & -0.087$\pm$0.057 & 0.05 & 1.28 \\
COMAi13004.035p28030.80 &  195.01682 & 28.00854 & 18.52 & 0 & 0.0213 & -0.009$\pm$0.020 & 0.09 & 0.90 \\
COMAi13005.347p275628.94 &  195.02229 & 27.94140 & 18.71 & 2 & 0.0237 & -0.094$\pm$0.026 & 0.10 & 1.84 \\
COMAi13005.403p28128.24 &  195.02249 & 28.02452 & 14.97 & 0 & 0.0197 & -0.047$\pm$0.000 & 0.04 & 2.18 \\
COMAi13005.684p275535.20 &  195.02367 & 27.92647 & 17.27 & 0 & 0.0265 & 0.100$\pm$0.003 & 0.38 & 0.78 \\
COMAi13005.755p28212.04 &  195.02399 & 28.03669 & 19.74 & 0 & 0.0194 & 0.016$\pm$0.010 & 0.16 & 1.02 \\
COMAi13006.399p28015.86 &  195.02663 & 28.00434 & 14.14 & 0 & 0.0241 & -0.112$\pm$0.003 & 0.06 & 1.73 \\
COMAi13007.123p275551.49 &  195.02968 & 27.93100 & 17.33 & 0 & 0.0259 & -0.196$\pm$0.006 & 0.03 & 1.94 \\
COMAi13008.002p28442.78 &  195.03329 & 28.07860 & 14.05 & 0 & 0.0241 & -0.136$\pm$0.016 & 0.24 & 0.53 \\
COMAi13011.143p28354.92 &  195.04639 & 28.06524 & 16.26 & 0 & 0.0244 & -0.074$\pm$0.032 & 0.12 & 0.07 \\
COMAi13012.864p28431.83 &  195.05357 & 28.07549 & 14.54 & 0 & 0.0250 & -0.074$\pm$0.035 & 0.36 & 0.05 \\
COMAi13013.808p28243.62 &  195.05751 & 28.04546 & 18.88 & 0 & 0.0298 & -0.081$\pm$0.046 & 0.05 & 0.71 \\
COMAi13014.170p28407.29 &  195.05902 & 28.06870 & 19.01 & 2 &  ...  & -0.092$\pm$0.013 & 0.35 & 1.53 \\
COMAi13014.745p28228.71 &  195.06142 & 28.04125 & 13.79 & 0 & 0.0191 & -0.099$\pm$0.004 & 0.18 & 2.24 \\
COMAi13015.725p28551.33 &  195.06552 & 28.09762 & 19.44 & 0 & 0.0172 & 0.010$\pm$0.010 & 0.33 & 2.96 \\
COMAi13016.364p275522.04 &  195.06816 & 27.92280 & 19.64 & 0 & 0.0149 & -0.018$\pm$0.016 & 0.04 & 0.95 \\
COMAi13016.521p275803.13 &  195.06883 & 27.96754 & 14.17 & 0 & 0.0154 & -0.091$\pm$0.001 & 0.18 & 0.17 \\
COMAi13016.675p275638.91 &  195.06947 & 27.94415 & 18.88 & 0 & 0.0182 & -0.089$\pm$0.019 & 0.37 & 0.86 \\
COMAi13016.964p275416.12 &  195.07065 & 27.90448 & 19.52 & 0 & 0.0303 & -0.018$\pm$0.006 & 0.30 & 2.70 \\
COMAi13017.618p275927.59 &  195.07343 & 27.99099 & 19.96 & 2 &  ...  & -0.017$\pm$0.042 & 0.11 & 1.82 \\
COMAi13017.643p275915.26 &  195.07352 & 27.98757 & 17.63 & 0 & 0.0199 & -0.076$\pm$0.002 & 0.26 & 0.31 \\
COMAi13018.349p28333.28 &  195.07646 & 28.05929 & 16.89 & 0 & 0.0333 & 0.201$\pm$0.004 & 0.20 & 0.66 \\
COMAi13018.543p28549.48 &  195.07727 & 28.09712 & 16.78 & 0 & 0.0260 & -0.046$\pm$0.006 & 0.17 & 1.32 \\
COMAi13018.715p275512.63 &  195.07797 & 27.92018 & 18.89 & 0 & 0.0231 & -0.015$\pm$0.025 & 0.09 & 2.43 \\
COMAi13018.782p275613.47 &  195.07826 & 27.93708 & 15.01 & 0 & 0.0171 & -0.071$\pm$0.001 & 0.63 & 1.71 \\
COMAi13018.873p28033.38 &  195.07878 & 28.00935 & 16.45 & 0 & 0.0207 & -0.143$\pm$0.003 & 0.14 & 0.27 \\
COMAi13020.399p28414.03 &  195.08502 & 28.07062 & 18.90 & 0 & 0.0235 & 0.005$\pm$0.006 & 0.09 & 1.94 \\
COMAi13021.672p275354.80 &  195.09030 & 27.89855 & 16.15 & 0 & 0.0162 & -0.070$\pm$0.003 & 0.17 & 0.79 \\
COMAi13022.156p28249.08 &  195.09236 & 28.04701 & 13.77 & 0 & 0.0272 & -0.097$\pm$0.003 & 0.17 & 0.30 \\
COMAi13022.656p275754.87 &  195.09439 & 27.96525 & 18.24 & 0 & 0.0214 & -0.060$\pm$0.032 & 0.03 & 1.64 \\
COMAi13022.949p275515.23 &  195.09564 & 27.92089 & 19.48 & 0 & 0.0176 & -0.047$\pm$0.019 & 0.31 & 0.96 \\
COMAi13023.476p28301.73 &  195.09785 & 28.05051 & 18.09 & 0 & 0.0235 & 0.017$\pm$0.076 & 0.10 & 0.17 \\
COMAi13024.823p275535.89 &  195.10347 & 27.92663 & 16.43 & 0 & 0.0264 & -0.046$\pm$0.001 & 0.31 & 0.59 \\
COMAi13025.977p28344.68 &  195.10825 & 28.06242 & 17.98 & 0 & 0.0250 & -0.055$\pm$0.072 & 0.03 & 0.14 \\
COMAi13026.152p28032.02 &  195.10904 & 28.00891 & 17.66 & 0 & 0.0185 & -0.054$\pm$0.015 & 0.16 & 0.76 \\
COMAi13027.339p28033.40 &  195.11398 & 28.00929 & 18.80 & 0 & 0.0213 & -0.106$\pm$0.010 & 0.08 & 0.34 \\
COMAi13027.569p28322.85 &  195.11485 & 28.05634 & 18.49 & 0 & 0.0193 & -0.041$\pm$0.008 & 0.24 & 0.35 \\
COMAi13027.608p28323.90 &  195.11504 & 28.05665 & 18.52 & 0 & 0.0193 & 0.001$\pm$0.015 & 0.29 & 0.24 \\
COMAi13027.879p275916.53 &  195.11618 & 27.98792 & 19.13 & 0 & 0.0320 & -0.057$\pm$0.022 & 0.23 & 0.42 \\
COMAi13027.971p275721.54 &  195.11655 & 27.95599 & 14.25 & 0 & 0.0233 & -0.067$\pm$0.000 & 0.18 & 0.43 \\
COMAi13028.376p275820.51 &  195.11824 & 27.97237 & 14.28 & 0 & 0.0257 & -0.027$\pm$0.006 & 0.38 & 0.40 \\
COMAi13030.949p28630.18 &  195.12897 & 28.10839 & 17.13 & 0 & 0.0172 & -0.059$\pm$0.008 & 0.41 & 0.41 \\
COMAi13031.949p275711.26 &  195.13307 & 27.95320 & 19.84 & 0 & 0.0196 & -0.013$\pm$0.018 & 0.17 & 1.74 \\
COMAi13032.493p275833.34 &  195.13535 & 27.97599 & 18.87 & 0 & 0.0286 & -0.011$\pm$0.004 & 0.38 & 0.88 \\
COMAi13032.520p28201.49 &  195.13548 & 28.03371 & 19.31 & 0 & 0.0212 & 0.034$\pm$0.033 & 0.04 & 1.85 \\
COMAi13032.613p28331.28 &  195.13588 & 28.05871 & 18.12 & 0 & 0.0246 & -0.045$\pm$0.028 & 0.09 & 0.88 \\
COMAi13032.960p275406.71 &  195.13728 & 27.90189 & 18.44 & 2 & 0.0222 & 0.005$\pm$0.076 & 0.11 & 0.98 \\
COMAi13033.335p275849.34 &  195.13885 & 27.98043 & 18.00 & 0 & 0.0172 & -0.040$\pm$0.022 & 0.03 & 0.56 \\
COMAi13034.427p275604.97 &  195.14346 & 27.93473 & 16.60 & 0 & 0.0291 & -0.016$\pm$0.008 & 0.13 & 1.19 \\
COMAi13035.418p275634.05 &  195.14761 & 27.94280 & 16.74 & 0 & 0.0231 & -0.054$\pm$0.002 & 0.45 & 0.36 \\
COMAi13035.990p275505.46 &  195.14995 & 27.91820 & 19.79 & 0 & 0.0232 & 0.018$\pm$0.031 & 0.25 & 1.61 \\
COMAi13036.581p275552.21 &  195.15244 & 27.93118 & 18.36 & 0 & 0.0197 & -0.129$\pm$0.016 & 0.20 & 1.75 \\
COMAi13036.669p275427.48 &  195.15277 & 27.90765 & 17.61 & 2 & 0.0207 & -0.135$\pm$0.020 & 0.14 & 0.51 \\
COMAi13037.299p275441.08 &  195.15541 & 27.91142 & 18.83 & 0 & 0.0203 & -0.066$\pm$0.026 & 0.02 & 0.80 \\
COMAi13038.731p28052.22 &  195.16139 & 28.01449 & 14.21 & 0 & 0.0254 & -0.034$\pm$0.002 & 0.25 & 1.51 \\
COMAi13039.754p275526.26 &  195.16566 & 27.92397 & 13.29 & 0 & 0.0250 & -0.093$\pm$0.001 & 0.18 & 1.71 \\
COMAi13040.853p275947.86 &  195.17021 & 27.99662 & 13.97 & 0 & 0.0236 & -0.045$\pm$0.004 & 0.31 & 0.15 \\
COMAi13041.192p28242.38 &  195.17162 & 28.04510 & 16.76 & 0 & 0.0287 & -0.047$\pm$0.003 & 0.22 & 0.15 \\
COMAi13042.519p28325.45 &  195.17719 & 28.05707 & 18.75 & 0 & 0.0195 & -0.153$\pm$0.013 & 0.05 & 3.36 \\
COMAi13042.833p275746.98 &  195.17847 & 27.96303 & 14.11 & 0 & 0.0280 & -0.060$\pm$0.001 & 0.20 & 2.11 \\
COMAi13042.889p28313.73 &  195.17868 & 28.05382 & 18.02 & 0 & 0.0226 & -0.051$\pm$0.010 & 0.33 & 0.71 \\
COMAi13044.127p28215.37 &  195.18385 & 28.03760 & 18.46 & 0 & 0.0292 & -0.005$\pm$0.020 & 0.17 & 2.01 \\
COMAi13044.634p28602.30 &  195.18599 & 28.10064 & 14.89 & 0 & 0.0219 & -0.089$\pm$0.000 & 0.48 & 0.14 \\
COMAi13048.045p28557.42 &  195.20018 & 28.09928 & 18.08 & 0 & 0.0208 & -0.012$\pm$0.051 & 0.10 & 0.98 \\
COMAi13048.646p28526.74 &  195.20268 & 28.09075 & 13.43 & 0 & 0.0230 & -0.117$\pm$0.007 & 0.25 & 1.71 \\
\hline
\caption{For each galaxy, Coma ID (see Paper II), right ascension and declination are given, together with S\'ersic magnitude, membership ratio, redshift,  colour gradient in the outer parts, ellipticity used in the Galphot fit and the reduced $\chi^2$ from the fit of the gradient to the outer part of the colour profile. }\label{table:sample}
\end{longtable}
\twocolumn

\begin{table}
\begin{tabular}{ll}
\hline
COMAi13017.020p28350.10 & late type, bar flattens gradient \\
COMAi13018.349p28333.28 & gradient in sky, maybe positive gradient\\
COMAi13011.807p28502.62 & too low S/N \\
COMAi13039.113p28035.46 & irregular galaxy \\
COMAi13042.508p28324.94 & in edge of a frame \\
COMAi13037.589p28059.94 & too faint (F814W = 19.9) \\
COMAi125945.531p28312.61 & in edge of a frame \\
COMAi13013.398p28311.81 & completely edge on\\
COMAi125914.447p28216.87 & too faint (F814W = 19.8) \\
COMAi13042.753p275816.88 & companions + bar + dust \\
COMAi13017.685p275718.92 & N4898a (overlapping)\\
COMAi13018.094p275723.51 & N4898b \\
COMAi13024.849p275921.86 & in low S/N gap between chips\\
COMAi125959.071p275841.26 & too low S/N \\
COMAi125944.776p275807.13 & too low S/N \\
COMAi125930.832p275810.32 & strong gradient in sky \\
COMAi125942.931p275954.39 & sky very uncertain\\
COMAi125956.707p275548.62 & overlapping, spiral arms\\
COMAi125950.703p275514.94 & strong gradient in sky\\
COMAi125944.825p275536.87 & too faint (F814W=19.99)\\
COMAi125923.730p275042.66 & in edge\\
COMAi125814.969p272744.81 & late type\\
COMAi125833.136p272151.77 & late type\\
COMAi125831.666p272342.04 & late type\\
COMAi125837.251p271035.59 & late type\\
COMAi125825.308p271200.04 & late type \\
COMAi125634.648p271339.03 & late type \\
COMAi125623.788p271402.30 & late type \\
COMAi125638.099p271304.09 & late type \\
COMAi125857.453p274706.28 & in edge \\
COMAi125854.632p274742.16 & in edge \\
COMAi125842.637p274537.86 & late type\\
COMAi125844.579p274458.30 & overlapping \\
COMAi125845.297p274650.75 & overlapping \\
\end{tabular}
\caption{Galaxies excluded from our sample.}\label{table:excluded}
\end{table}

\clearpage
\begin{table}
\begin{tabular}{|l|r|r|l|r|r|r|r|}
\hline
  \multicolumn{1}{|c|}{Galaxy} &
  \multicolumn{1}{c|}{$\Delta$age} &
  \multicolumn{1}{c|}{$\Delta$[M/H]} &
  \multicolumn{1}{c|}{m$_B$$^1$} &
  \multicolumn{1}{c|}{$\Delta$Mgb} &
  \multicolumn{1}{c|}{$\Delta$col} \\
\hline
  NGC4673 & -0.1593 $\pm$0.096 & -0.1049 $\pm$0.1239 & 13.7 & -0.023 $\pm$0.003 & -0.15 $\pm$0.019\\
  NGC4692 & 0.163 $\pm$0.1103 & -0.3817 $\pm$0.1285 & 14.0 & -0.014 $\pm$0.004 & -0.09 $\pm$0.026\\
  NGC4839 & 0.28 $\pm$0.1708 & -0.3535 $\pm$0.154 & 13.0$^{(2)}$ & -0.016 $\pm$0.02 & -0.10 $\pm$0.129\\
  NGC4842A & -0.2147 $\pm$0.2617 & 0.0637 $\pm$0.2364 & 14.9 & - & -\\
  NGC4864 & 0.49 $\pm$0.1762 & -0.5792 $\pm$0.1618 & 14.8 & -0.016 $\pm$0.009 & -0.10 $\pm$0.058\\
  NGC4865 & 0.0672 $\pm$0.1232 & -0.2587 $\pm$0.1416 & 14.6 & -0.026 $\pm$0.01 & -0.17 $\pm$0.065\\
  NGC4874 & 0.5845 $\pm$0.1029 & -0.7181 $\pm$0.1382 & 13.7 & -0.044 $\pm$0.01 & -0.28 $\pm$0.065\\
  NGC4875 & 0.1777 $\pm$0.139 & -0.3231 $\pm$0.2141 & 15.6$^{(3)}$ & - &  -\\
  NGC4889 & 0.2578 $\pm$0.0711 & -0.4363 $\pm$0.112 & 13.0 & -0.012 $\pm$0.005 & -0.077 $\pm$0.032\\
  NGC4908 & 0.0131 $\pm$0.2231 & -0.2354 $\pm$0.2812 & 14.9 & -0.004 $\pm$0.005 & -0.026 $\pm$0.032\\
  IC832 & -0.1432 $\pm$0.1404 & 0.0902 $\pm$0.1483 & 15.0$^{(3)}$ & -0.003 $\pm$0.004 & -0.019 $\pm$0.026\\
  IC3957 & 0.2298 $\pm$0.2385 & -0.4788 $\pm$0.2636 & 15.6$^{(3)}$ & -0.006 $\pm$0.009 & -0.039 $\pm$0.058\\
  IC3959 & 0.0731 $\pm$0.1344 & -0.1642 $\pm$0.1655 & 15.2 & -0.009 $\pm$0.004 & -0.058 $\pm$0.026\\
  IC3963 & 0.336 $\pm$0.134 & -0.5703 $\pm$0.1788 & 15.7$^{(3)}$ & -0.011 $\pm$0.008 & -0.071 $\pm$0.052\\
  IC3973 & 0.4122 $\pm$0.1776 & -0.3849 $\pm$0.1643 & 15.2 & -0.005 $\pm$0.004 & -0.0323 $\pm$0.026\\
  IC4042 & -0.1029 $\pm$0.4661 & -0.3628 $\pm$0.6512 & 15.4 & -0.05 $\pm$0.007 & -0.323 $\pm$0.045\\
  IC4051 & -0.0582 $\pm$0.0852 & -0.3693 $\pm$0.1261 & 14.8 & -0.03 $\pm$0.014 & -0.194 $\pm$0.090\\
  CGCG159-41 & 0.1629 $\pm$0.5742 & -0.0679 $\pm$0.6968 & 15.5  & -0.022 $\pm$0.005 & -0.142 $\pm$0.032\\
  CGCG159-43 & 0.5593 $\pm$0.1614 & -0.6726 $\pm$0.1936 & 15.3 & -0.014 $\pm$0.004 & -0.090 $\pm$0.026\\
  CGCG159-83 & -0.4099 $\pm$0.131 & 0.0939 $\pm$0.1566 & 14.9 & -0.023 $\pm$0.008 & -0.148 $\pm$0.052\\
  CGCG159-89 & -0.0815 $\pm$0.141 & -0.3979 $\pm$0.1969 & 14.8  & -0.016 $\pm$0.004 & -0.103 $\pm$0.026\\
\hline\end{tabular}
\caption{Data from Sanchez-Blazquez et al. extended with own data. Columns show age, metallicity and Mg b gradients, with formal errors. B-band magnitudes are all from (1) \citet{FalKurGel99} except (2) \citet{GilBoiMad07}  and (3) \citet{deVdeVCor91}}
\end{table}

\clearpage
\begin{figure*}
\centering
\begin{tabular}{cc}
\scalebox{0.4}[0.4]{\includegraphics{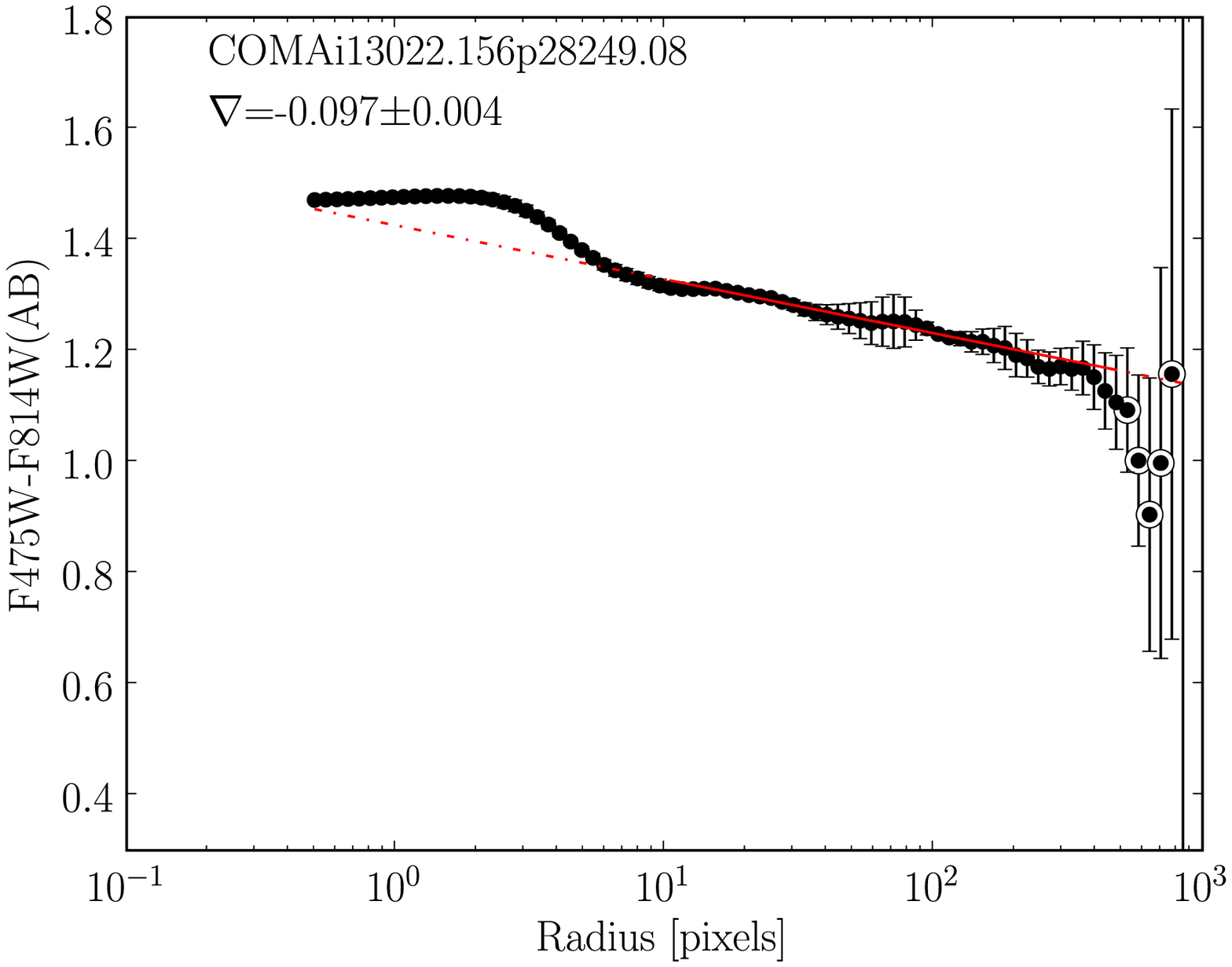}} & 
\scalebox{0.4}[0.4]{\includegraphics{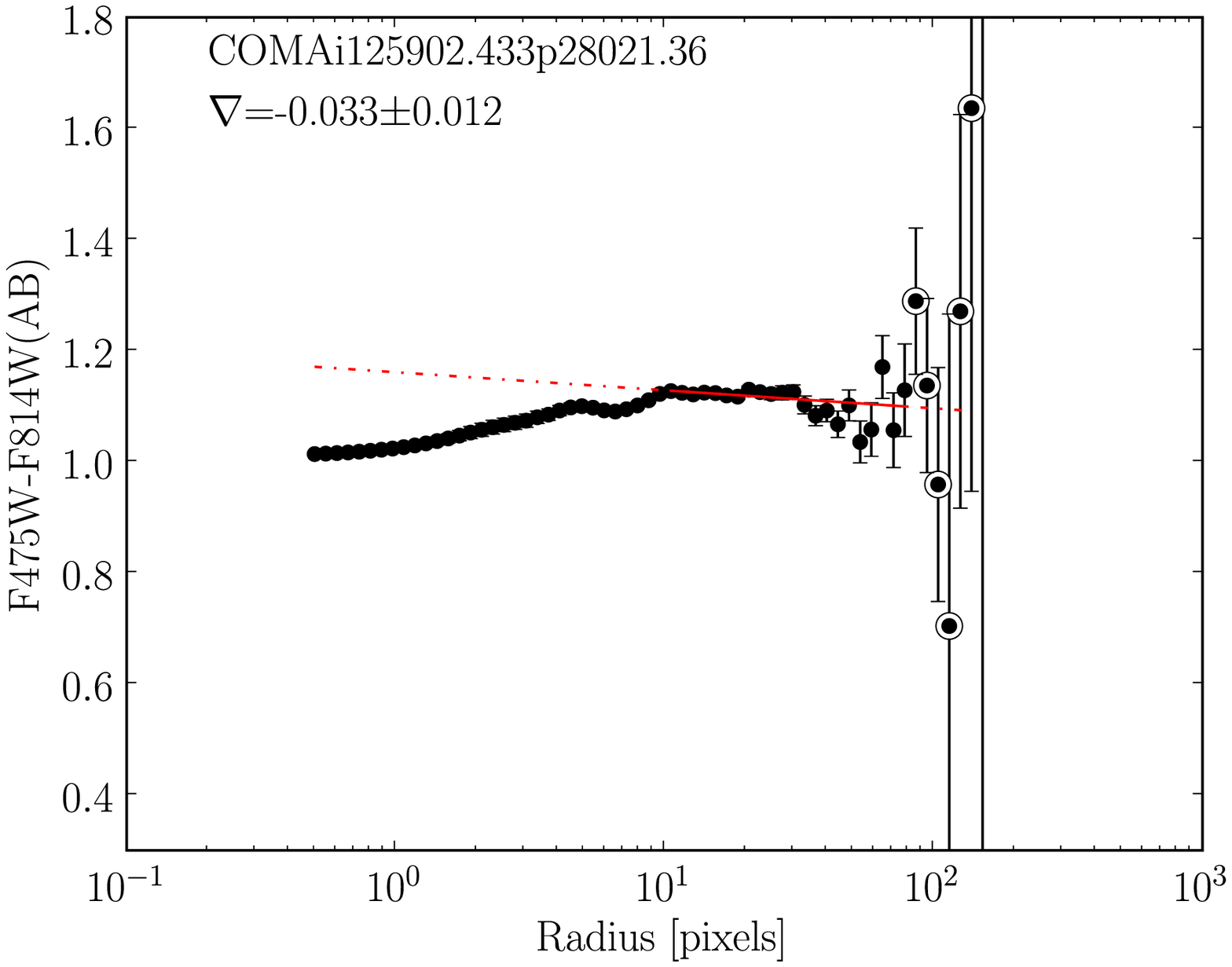}} \\ 
\scalebox{0.4}[0.4]{\includegraphics{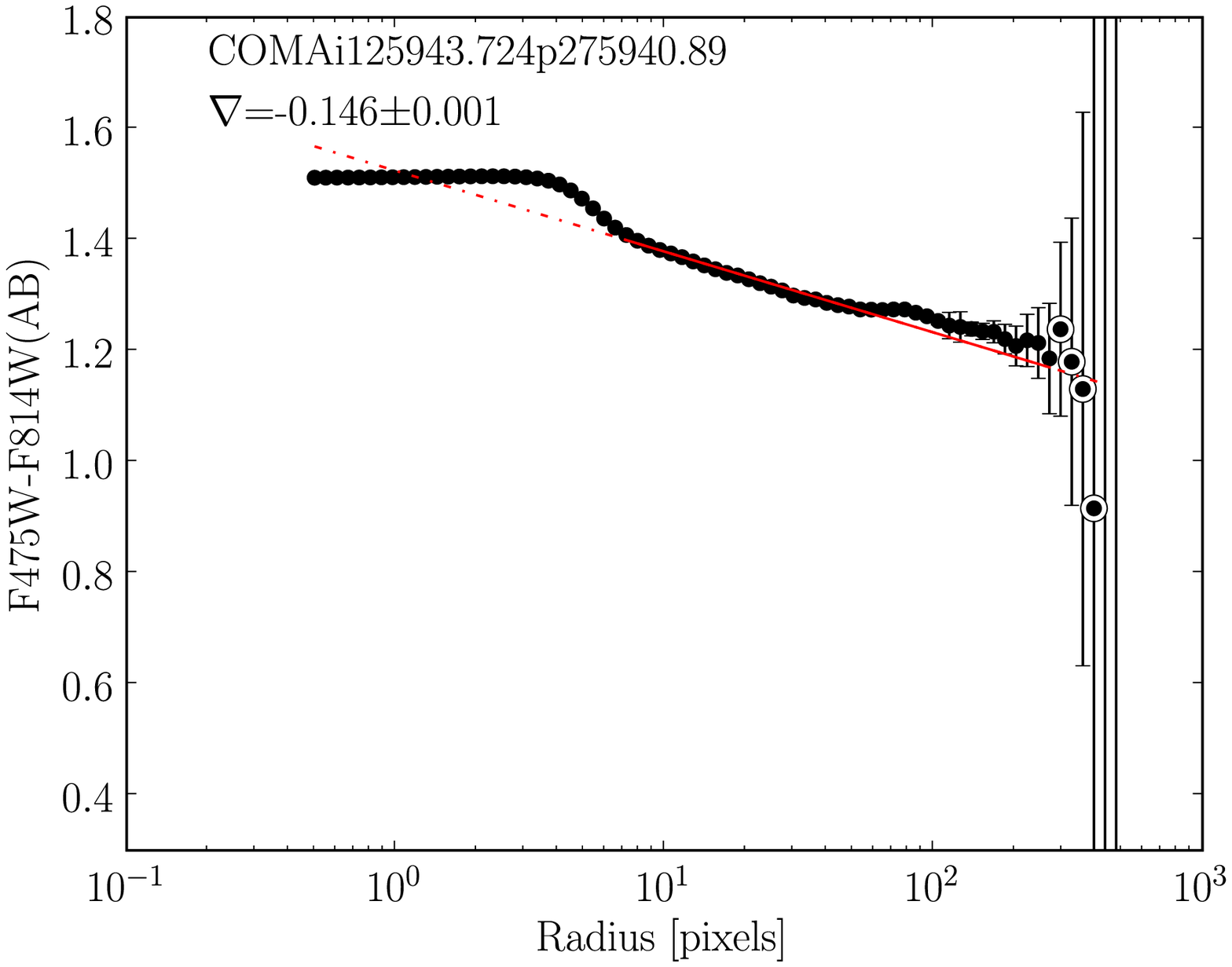}} & 
\scalebox{0.4}[0.4]{\includegraphics{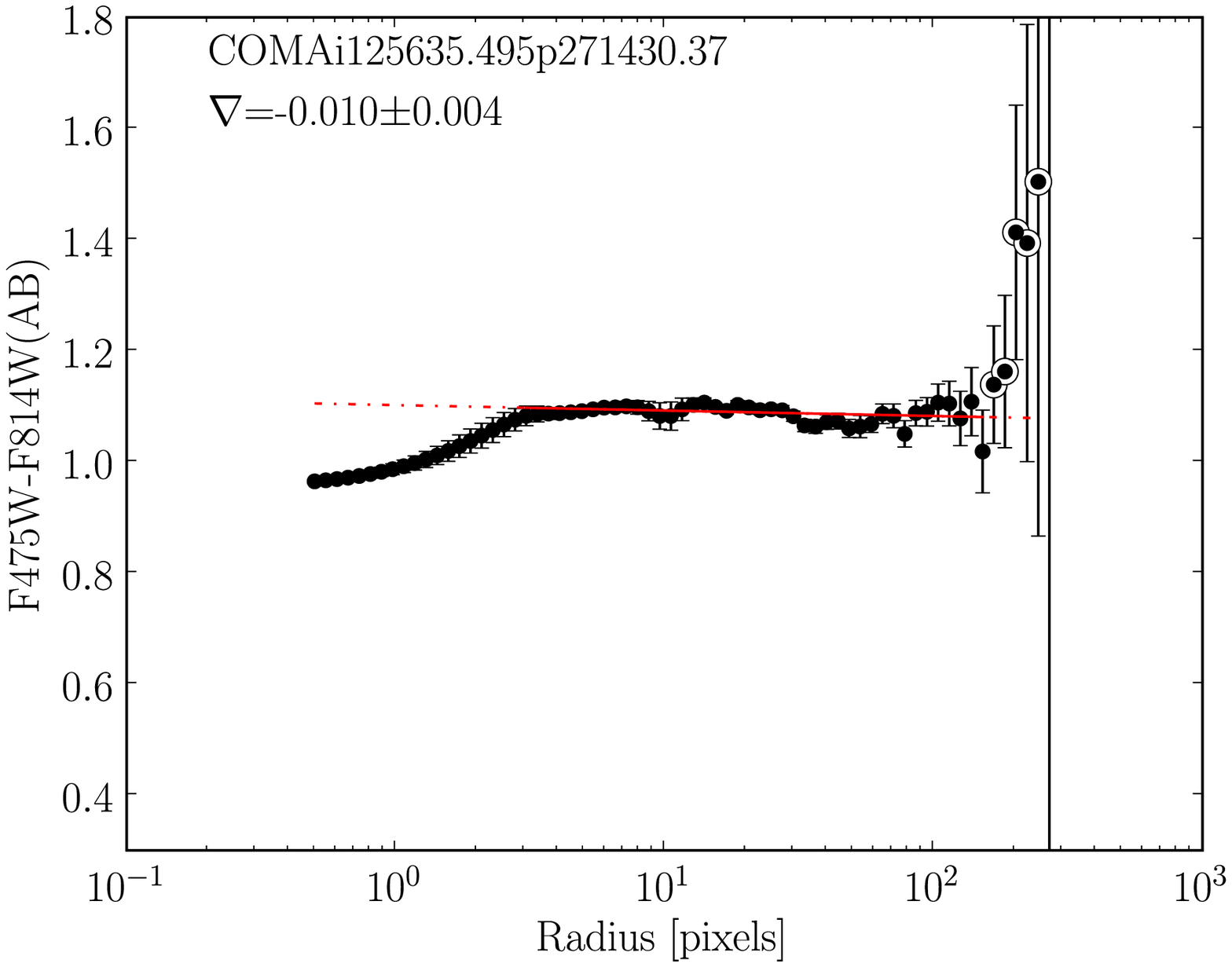}}\\
\scalebox{0.4}[0.4]{\includegraphics{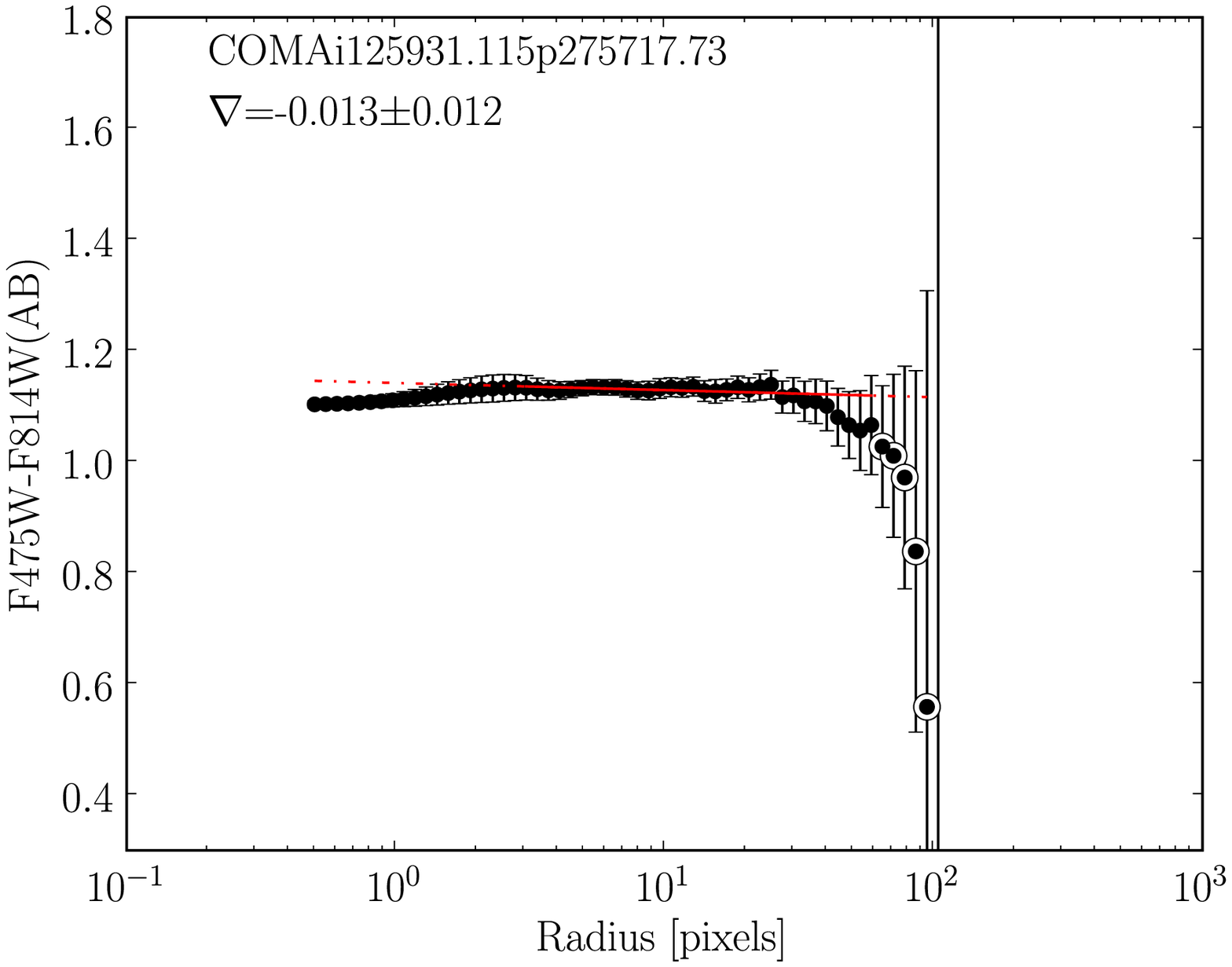}}&
\scalebox{0.4}[0.4]{\includegraphics{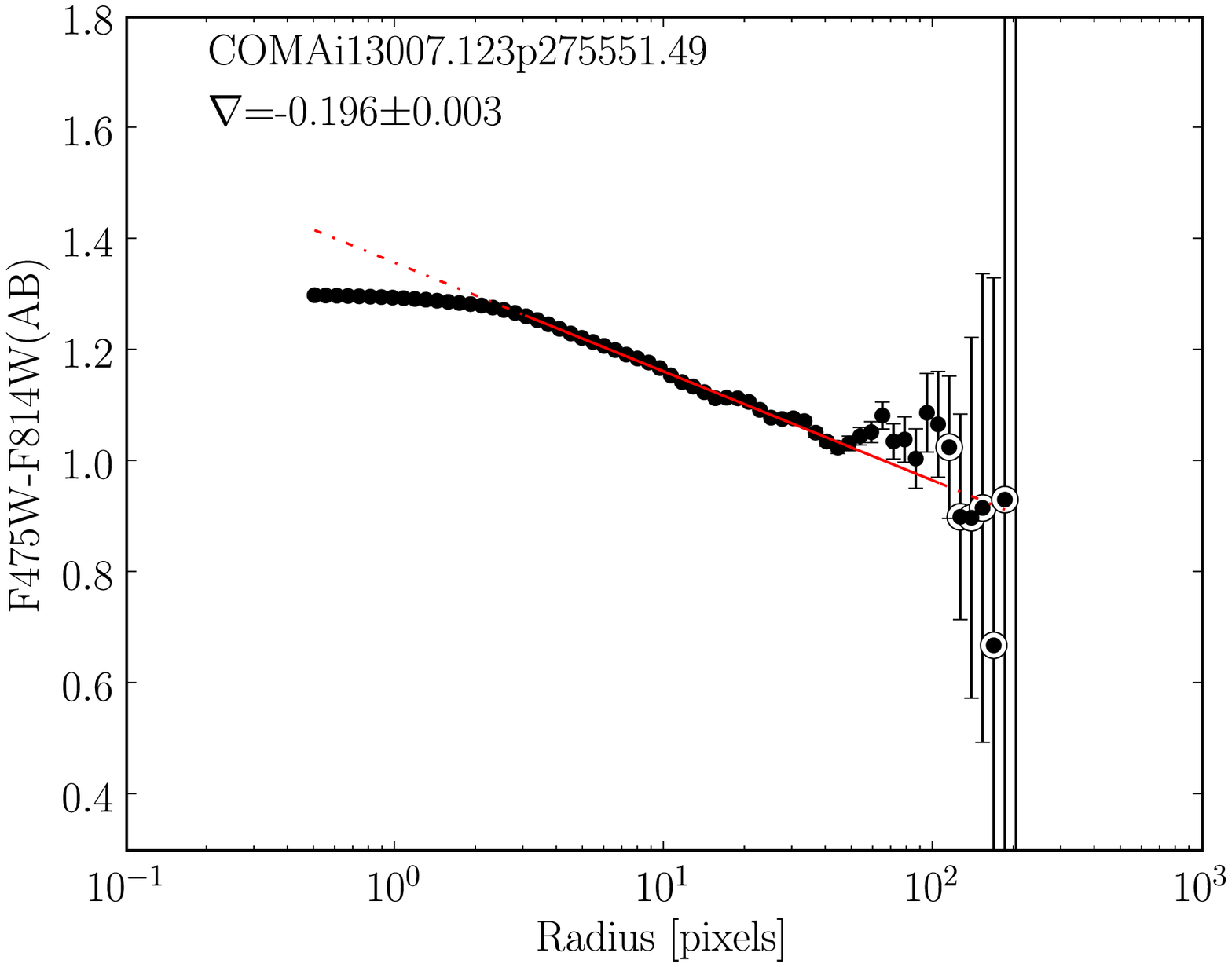}}\\
\end{tabular}
\caption{Colour profiles for a subset of our galaxies, sorted by magnitude in anticlockwise direction. Brightest galaxy here is \texttt{COMAi13022.156p28249.08} (F814W(AB)=13.77) and faintest galaxy is \texttt{COMAi125902.433p28021.36} (F814W(AB)=18.93). \texttt{COMAi13022.156p28249.08} is a barred galaxy (which is why the errors increase and decrease over the profile) and has a central dust disk.Also \texttt{COMAi25943.724p275940.89} has a central dust disk, which is seen face on. \texttt{COMAi125902.433p28021.36} and \texttt{COMAi125635.495p271490.37} are typical examples of dwarf galaxies, whereas \texttt{COMAi13007.123p275551.49} is an example of a faint system with a steep gradient and high S\'ersic index. The profiles of the full sample can be found in Appendix B of the electronic edition of this paper.}
\end{figure*}

\end{document}